\documentclass[runningheads]{llncs}
\usepackage[T1]{fontenc}
\usepackage{graphicx}
\usepackage{tabularx}        
\usepackage{makecell}        
\usepackage{multirow}
\usepackage{bm}
\usepackage{adjustbox}
\usepackage{amssymb}
\usepackage{caption}
\usepackage{subcaption}
\usepackage{array}
\usepackage{makecell}
\usepackage{doi}
\usepackage[strings]{underscore}
\usepackage{amsmath}
\AtBeginEnvironment{table}{\setlength\belowcaptionskip{0pt}}
\usepackage{enumerate}
\usepackage[shortlabels]{enumitem}
\usepackage{xspace}
\usepackage{amsfonts}
\usepackage{dsfont}

\usepackage{mathrsfs}

\usepackage{anyfontsize}
\usepackage{lmodern}
\DeclareMathAlphabet{\mathpzc}{OT1}{pzc}{m}{it}
\usepackage{color}
\usepackage[dvipsnames]{xcolor}
\usepackage{hyperref}
\hypersetup{
    colorlinks=true,
    linkcolor=NavyBlue,
    urlcolor=NavyBlue,
    linktoc=all,
    citecolor=NavyBlue}

\definecolor{blue3}{RGB}{0,76,153}
\newcommand{\ka}[1]{{\color{blue3}\texttt#1}}
\usepackage{notoccite}

\begin{document}
\title{On the Formalization of Network Topology Matrices in HOL}
%
%

\author{Kubra Aksoy$^{1}$, Adnan Rashid$^{1,2}$, Osman Hasan$^{2}$ and Sofi\`{e}ne Tahar$^{1}$}
%
%
%
\institute{$^{1}$Department of Electrical and Computer Engineering,\\
Concordia University, Montreal, QC, Canada \\
\email{\{k\_aksoy, rashid, tahar\}@ece.concordia.ca}\\[2em]
$^{2}$School of Electrical Engineering and Computer Science\\
National University of Sciences and Technology, Islamabad, Pakistan\\
\email{{osman.hasan@seecs.nust.edu.pk}}} 

\authorrunning{K. Aksoy et al.}

%
\maketitle              
\begin{abstract}

Network topology matrices are algebraic representations of graphs that are widely used in modeling and analysis of various applications including electrical circuits, communication networks and transportation systems. In this paper, we propose to use Higher-Order-Logic (HOL) based interactive theorem proving to formalize network topology matrices. In particular, we formalize adjacency, degree, Laplacian and incidence matrices in the Isabelle/HOL proof assistant. Our formalization is based on modelling systems as networks using the notion of directed graphs (unweighted and weighted), where nodes act as components of the system and weighted edges capture the interconnection between them. Then, we formally verify various classical properties of these matrices, such as indexing and degree. We also prove the relationships between these matrices in order to provide a comprehensive formal reasoning support for analyzing systems modeled using network topology matrices. To illustrate the effectiveness of the proposed approach, we formally analyze the Kron reduction of the Laplacian matrix and verify the total power dissipation in a generic resistive electrical network, both commonly used in power flow analysis.

\end{abstract}
\section{Introduction}
\label{Sec:Intro}

Network systems are interconnected structures widely used to model and analyze real-world problems encountered in various domains including biology, sociology, computer science, mathematics and engineering \cite{newman2018networks}. The foundational purpose of networks is to facilitate the transmission of physical entities (e.g., electricity, water and mass) and information (e.g., digital signals and gene regulatory) according to systems' governing laws, such as conservation laws \cite{b4} or network protocols \cite{karl2007protocols}. The analysis of these networks involves an examination of the structural properties, commonly known as network topology, which helps to understand the behavior of networks based on the arrangement of these entities and their interrelationships (physical or logical).
\par In network topology, graphs are employed to construct the mathematical foundation for the visualization of different components, which vary from one system to another. For instance, electrical networks~\cite{b4} can be modeled as systems comprising interconnected elements, such as resistors, conductors, inductors and current/voltage sources. In the corresponding graph representation, these elements are depicted as edges, while their connection points are represented by nodes. In addition, weights can be assigned to the edges to reflect specific information regarding these elements, such as values of resistors, capacitors, etc. Similarly, transportation systems~\cite{cascetta2009transportation}, such as airport and road networks, can be captured using the notion of graphs, where terminals or cities are represented by nodes and their interconnection through routes are modelled as edges, often incorporating additional information, such as cost and distance for each connection, represented by weights of the edges.\\
\indent The topology analysis of network systems often requires establishing a connection between graphs and matrices. Matrices~\cite{b6}, as compact algebraic representations, provide a powerful framework for efficiently analyzing network systems. This algebraic formulation allows us to leverage linear algebraic methods, such as Gauss elimination, Kron reduction and matrix operations~\cite{b6}, especially when dealing with complex and large-scale systems. There are several network topology matrices~\cite{bullo2018lectures}, such as the adjacency, the incidence, the loop, the degree and the Laplacian matrices, that can be derived from graphs to capture various structural and fundamental properties of the systems.\\
\indent In the analysis of electrical networks, Kirchhoff was among the first to employ topology matrices to formulate Kirchhoff's Current Law (KCL) and Kirchhoff's Voltage Law (KVL) \cite{kirchhoff1958solution}. Similarly, the topological analysis of multibody systems also relies on using topology matrices, which greatly facilitate in understanding their kinematic behavior~\cite{b2}. In mathematical chemistry, Laplacian matrices are used to analyze complex molecular graphs and to determine their topological indexes~\cite{merris1994laplacian}. Motivated by the widespread use of such matrices across diverse applications, particularly in safety-critical domains, it is essential to establish a rigorous formal foundation for this field.\\
\indent Mathematically, the \textit{adjacency matrix}~\cite{b1} encodes information about the connectivity between nodes in a network, representing the presence or absence of edges between every pair of nodes. The \textit{degree matrices} of weighted graphs are characterized by the in (out) degree of each node, which is equal to the total number of incoming (outgoing) weights of edges relative to the node. Similarly, the \textit{incidence matrix} provides a relationship between the nodes and edges of a network system~\cite{b1}, while the \textit{loop matrix} presents a relationship between loops and edges~\cite{b1}. On the other hand, the \textit{Laplacian matrix}, often associated with weighted directed graphs, captures the overall effect of the interaction between nodes, incorporating their connectivity and weights of the associated edges. Moreover, a Laplacian matrix can be constructed using adjacency and degree matrices, which are also associated with the incidence matrices. While there are other topological matrices that can represent different types of graphs, our focus in this paper is primarily on the Laplacian matrix and its closely related topological matrices.\\
\indent Conventionally, these network topology matrices have been analyzed using paper-and-pencil proofs and computer-based simulation approaches. However, the former tends to be error-prone, especially for large and complex systems. As discussed in~\cite{qadir2014applying}, computer-based simulation methods do not provide sound guarantees, as they typically rely on trial-and-error processes and may skip some very safety-critical cases. Moreover, their accuracy cannot be ensured due to the presence of unverified algorithms existing within the underlying computational tools. In contrast, formal methods, particularly interactive theorem proving,  offer a rigorous, logic-based framework for the computer-assisted mathematical modeling and analysis. These approaches support deductive reasoning grounded in first- or higher-order logics. In this context, HOL-based interactive theorem provers are particularly well-suited for analyzing network topology matrices, offering high expressiveness and rich mathematical libraries that facilitate precise and trustworthy analysis.\\
\indent In this paper, we use the Isabelle/HOL proof assistant to formalize network topology matrices. We begin by developing a formal network topology system using a modular approach in Isabelle/HOL. This system extends our previous work~\cite{aksoy2025faecnttp} by systematically declaring new locales that describe different classes, including weighted and symmetric networks, inheriting from the existing locales. A key contribution of this paper is the formalization of the adjacency, degree and Laplacian matrices derived from the network modeled as a weighted directed graph, together with rigorous proofs of their fundamental properties such as degree and indexing relations. Furthermore, we introduce and formalize both the in-incidence and out-incidence matrices and establish their relationships with the incidence matrix previously defined in~\cite{aksoy2025faecnttp}. Collectively, these extensions enable the formal verification of the interrelationships among all these matrices, ensuring the consistency and soundness of our framework. To demonstrate the practical applicability of our proposed approach, we apply our formalization to two major structures: undirected weighted graphs and weighted directed graphs. Specifically, we formally analyze the Kron reduction~\cite{dorfler2012kron} of the Laplacian matrix for weighted directed graphs. Kron reduction is a widely used algebraic method in power system analysis. We also formalize a generalized version of Ohm's law for resistive circuits and use it to formally verify total power dissipation using Laplacian matrices in Isabelle/HOL. The choice of Isabelle/HOL is motivated by its robust reasoning capabilities and comprehensive library support for matrices and graph theory. The complete Isabelle/HOL code for our formalization is publicly available at~\cite{b25}.\\
\indent The rest of the paper is organized as follows: We discuss related work in Section~\ref{SEC:rw}. Section~\ref{SEC:preliminaries} provides an overview of the Isabelle/HOL proof assistant and introduces some fundamental symbols/notations, definitions and lemmas of matrix theory in Isabelle/HOL that are necessary for understanding the rest of the paper. Section~\ref{sec4graph} presents the formalization of the network system, while Section~\ref{sec5adl} details the formalization of the adjacency, degree and Laplacian matrices, along with the verification of their classical properties. Section~\ref{sec6incirel} describes the formalization of the incidence matrices and explores their relationships with the previously formalized network topology matrices. Next, we provide two applications of the Laplacian matrices in Section~\ref{sec7app}. Section~\ref{SEC:Dissc}  discusses the experiences and challenges encountered during the formalization process. Finally, we conclude the paper in Section~\ref{SEC:Conclusion}.

\section{Related Work}
\label{SEC:rw}
In this section, we discuss the most relevant contributions regarding the formalization of networks, graphs and the topology matrices. For instance, Butler et al.~\cite{butler1998pvs} used PVS~\cite{PVS} to formalize a graph theory library with fundamental concepts based on directed graphs. Noschinski~\cite{noschinski2015graph} used Isabelle/HOL to develop a generic graph theory using record+locale structure, with a major focus on directed graph notions. Koutsoukou-Argyraki et al.~\cite{kosaian2023first} formalized the Balog–Szemerédi–Gowers theorem using a general library for undirected graphs in Isabelle/HOL. Similarly, Doczkal et al.~\cite{b18} formalized a graph library in the Coq/SSReflect theorem prover~\cite{ROCQ}. More recently, Narv{\'a}ez et al.~\cite{narvaez2024formalizing} used Lean 4~\cite{LEAN4} to formalize the finite Ramsey theorem that is built upon Mathlib's graph theory. However, these contributions mostly focus on generic formalizations of notions and theorems in graph theory.

Some contributions on the formalization of networks focus on graph algorithms and their verification of some network systems, without addressing network topology matrices. For instance, Wong~\cite{b16} formally analyzed the railway track networks based on simple graphs using HOL~\cite{HOL}. Similarly, Lee~\cite{lee2005correctnesss} used the Mizar theorem prover~\cite{MIZAR} to verify several graph algorithms, such as Prim, Dijkstra, etc. Diekmann et al.~\cite{b19} used Isabelle/HOL to formalize graphs and used it for formally verifying various network security policies. Similarly, Lammich et al.~\cite{lammich2019formalizing} formalized network flow algorithms in Isabelle/HOL by using their own formalization of graph theory and formally verified their correctness and time complexity. Building upon the work in~\cite{noschinski2015graph}, Kov\'acs et al.~\cite{kovacs2020formalizing} verified an algorithm to compute the longest strictly decreasing ordered graph trail. Similarly, Lochbihler~\cite{lochbihler2022mechanized} formally proved the maximum-flow minimum-cut theorem by formalizing finite networks with flows and cuts in Isabelle/HOL. More recently, Tekriwal et al.~\cite{tekriwal2024formally} formalized the weighted-mean subsequence reduced (W-MSR) algorithm to address the problem of consensus in a network modeled by a directed graph in the Coq theorem prover. Whilst the aforementioned contributions are able to verify certain graphs and network related algorithms and formalize important mathematical concepts, they do not consider topology matrices in analyzing systems that require algebraic representations.

\par There have been a few efforts to formalize the network topology matrices in various interactive theorem provers. For example, Heras et al.~\cite{b20} used Coq/SSReflect to formalize undirected graphs and their coresponding incidence matrices to formally analyze 2D digital image processing systems. Similarly, Edmonds et al.~\cite{b21} used Isabelle/HOL to formalize incidence matrices of design, which are an algebraic representation of a more general combinatorial structure than  undirected graphs, and further used them to prove Fisher's inequality. 
In the Lean 4 mathlib library~\cite{leanmathlib}, basic formalizations of the adjacency, incidence and Laplacian matrices are available, but only for simple graphs. More recently, we developed a basic  formalization of incidence and loop matrices to formally analyze electrical circuit network topologies in Isabelle/HOL~\cite{aksoy2025faecnttp}. Unlike the work in~\cite{b21}, we formalized the incidence matrices for directed graphs by incorporating the direction aspect and considering the entries of the matrix as $0$, $1$ and $-1$. We also employed the formalization of loop matrices for the formal kinematic analysis of epicyclic bevel gear trains in Isabelle/HOL~\cite{aksoy2024formal}. However, to the best of our knowledge, none of these existing works provide a comprehensive formalization of the Laplacian,  adjacency, and degree matrices for networks represented as weighted directed graphs, which constitutes the primary focus of this paper.

\vspace*{-.2cm}
\section{Preliminaries} \label{SEC:preliminaries}
\vspace*{-.2cm}
In this section, we provide a brief overview of the Isabelle/HOL proof assistant and some definitions, lemmas from the existing matrix theory that are used in our formalization of the network topology matrices and are necessary for understanding the rest of the paper.
\vspace*{-.1cm}
\subsection{Isabelle/HOL Proof Assistant} 
\vspace*{-.1cm}
Isabelle/HOL~\cite{isabelle} is a higher-order-logic (HOL) based interactive proof assistant used for the formalization of mathematics (e.g.,~\cite{GodelIsa} and~\cite{GREENisabelle}) and the verification of algorithms (e.g.,~\cite{verifiedalg} and~\cite{ver2}), software (e.g.,~\cite{software}) and hardware systems (e.g.,~\cite{hardware}). The core of the tool relies on a small trusted kernel consisting of basic axioms and primitive inference rules. All lemmas and theorems are verified based on either these inference rules or already verified lemmas. An Isabelle/HOL theory consists of a collection of definitions, data types, functions and theorems. Isabelle includes the Isabelle/Isar proof language~\cite{isar}, which enables proofs to be written in a human-readable way. 

\par One of the key features of Isabelle/HOL is its integration with automation tools through \textit{Sledgehammer}~\cite{sled}, which offers the support of Automatic Theorem Provers (ATP) and Satisfiability-Modulo-Theories (SMT)~\cite{smt} solvers to prove statements of lemmas/theorems, automatically. Isabelle/HOL also has a rich collection of theories that are stored in the related Archive of Formal Proof (AFP). Table~\ref{tab:isanotions} provides some commonly used Isabelle/HOL symbols and functions that appear in the subsequent sections.

\par Isabelle/HOL also has a powerful module system of proof contexts called \textit{locales}, an extension of the Isar language. Locales~\cite{locales} can be used to modularly model different algebraic structures, which are based on \textit{contexts}, and their connections. A context specifies parameters and assumptions using the keywords \textit{fixes} and \textit{assumes}~\cite{locales}. Locales also support a hierarchical structure, allowing new contexts to be constructed by composing existing ones. Consequently, definitions and theorems formalized within one locale can be reused through locale inheritance, enhancing extensibility.
Moreover, locales can be manipulated using \textit{interpretation} command, e.g., it enables an instantiation of a generic locale with a specific parameters, which in turn may be of a specific type like \texttt{int} or \texttt{complex}. In this work, we adopt a locale-based approach to formalize network systems, which is presented in Section~\ref{sec4graph}.
\vspace{-0.3cm}
\begin{table}[htbp]%
    \centering%
    \begin{adjustbox}{width=\textwidth}
    \begin{tabular}{|c|c|}%
    \hline
    \textbf{Isabelle Symbols/Notions} & \textbf{Meaning} \\ \hline\hline
    $\bigwedge$         & Universal quantification in meta-logic                 \\ \hline
    $\Longrightarrow$           & Implication in meta-logic                   \\ \hline
    $\land$             & Logical and                   \\ \hline  
    [\:]       & Empty list                     \\ \hline
    {\fontfamily{cmtt}\selectfont $'$a ::\:x }           &              Generic data type variable belongs to type class \texttt{x}    \\ \hline
    
    $\lambda x. f$          & Function that maps $x$ to $f (x)$               \\ \hline
    {\fontfamily{cmtt}\selectfont es!i }         & i$^{th}$ element of the list  {\fontfamily{cmtt}\selectfont es}              \\ \hline
     {\fontfamily{cmtt}\selectfont length es}         & Function that computes the size of the list {\fontfamily{cmtt}\selectfont es}              \\ \hline
     \textup{{\fontfamily{cmtt}\selectfont A\textsuperscript{T}}}  &       Transpose of matrix {\fontfamily{cmtt}\selectfont A}  \\ \hline
       {\fontfamily{cmtt}\selectfont v} $\$$  {\fontfamily{cmtt}\selectfont i}           &              i$^{th}$ element of the vector {\fontfamily{cmtt}\selectfont v}  \\ \hline
      {\fontfamily{cmtt}\selectfont A} $\$\$$  {\fontfamily{cmtt}\selectfont (i,j)}     &   i$^{th}$ row and  j$^{th}$ column element of the matrix A 
      in JNF{*}           \\ \hline\hline
  
   \multicolumn{2}{r}{{\footnotesize *\textit{JNF} \textit{is an abbreviation of Jordan Normal Form library in Isabelle/HOL}}}\\ 
    \end{tabular}
    \end{adjustbox}
     \caption{Isabelle/HOL Symbols}%
     \label{tab:isanotions}
\end{table}
\vspace*{-1cm}
\subsection{Matrix Libraries in Isabelle/HOL}

There are several matrix libraries available in Isabelle/HOL, such as HOL-Analysis (HA) and Jordan Normal Form (JNF) matrix libraries. The theory of vectors and matrices is formalized by Chaieb et al.~\cite{b13} as a part of the HA library that was primarily inspired by the Harrison’s formalizations available in the HOL Light theorem prover~\cite{b11}. Similarly, the JNF matrix library developed by Thiemann et al.~\cite{b14}, which includes the foundation of block matrices for formally verifying JNF. Moreover, the JNF library provides many formally verified properties of matrix and linear algebra. While the HA library offers valuable functionality in many mathematical models, its current implementation presents challenges for the use of block matrices with varying dimension due to type restriction on row and column dimensions. This limitation can reduce its applicability for potential applications, such as spectral analysis of networks~\cite{bullo2018lectures}. Therefore, we utilize the JNF library to develop our proposed formalization.

We now present some of the common Isabelle/HOL functions that are used in the proposed formalization. All matrices in JNF are defined via the type $'${\fontfamily{cmtt}\selectfont a mat} in the form of a triple  {\fontfamily{lmtt}\selectfont (nr,nc,f)} that is formalized as follows~\cite{b14}: 
\vspace*{-0.05cm}
\begin{flushleft}
{\fontfamily{cmtt}\selectfont \small {\bf{typedef}} 
\textcolor{blue3}{$'$a\hspace*{-.1cm} mat}\hspace*{-.1cm} =\hspace*{-.1cm} $\{$(nr,nc,mk\textunderscore mat\hspace*{-.1cm} nr\hspace*{-.1cm} nc\hspace*{-.1cm} f)\hspace*{-.1cm} |\hspace*{-.1cm} nr\hspace*{-.1cm} nc\hspace*{-.1cm} f\hspace*{-.1cm} ::\hspace*{-.1cm} nat\hspace*{-.1cm} $\times$\hspace*{-.1cm} nat\hspace*{-.1cm} $\Rightarrow$\hspace*{-.1cm} $'$a.\hspace*{-.1cm} True$\}$}
\end{flushleft}
\vspace*{-0.05cm}
The following introduction rule is employed to prove matrix equality, which is frequently used in our proposed formalization. 
\vspace*{-0.05cm}
\begin{flushleft}
{\fontfamily{cmtt}\selectfont \small {\bf{lemma}} \ka{eq\textunderscore matI[intro]}:  \\
\hspace*{.005in} $\bigwedge$\hspace*{-.1cm} i\hspace*{-.1cm} j.\hspace*{-.1cm} i\hspace*{-.1cm} $<$\hspace*{-.1cm} dim\textunderscore row\hspace*{-.1cm} B\hspace*{-.1cm} $\Longrightarrow$\hspace*{-.1cm} j\hspace*{-.1cm} $<$\hspace*{-.1cm} dim\textunderscore col\hspace*{-.1cm} B\hspace*{-.1cm} $\Longrightarrow$\hspace*{-.1cm} A\hspace*{-.1cm} \$\$\hspace*{-.1cm} (i,j)\hspace*{-.1cm} =\hspace*{-.1cm} B\hspace*{-.1cm} \$\$\hspace*{-.1cm} (i,j)  \\ \hspace*{1.5cm} $\Longrightarrow$\hspace*{-.1cm} 
dim\textunderscore row\hspace*{-.1cm} A\hspace*{-.1cm} =\hspace*{-.1cm} dim\textunderscore row\hspace*{-.1cm} B\hspace*{-.1cm} $\Longrightarrow$\hspace*{-.1cm} dim\textunderscore col\hspace*{-.1cm} A\hspace*{-.1cm} =\hspace*{-.1cm} dim\textunderscore col\hspace*{-.1cm} B\hspace*{-.1cm} $\Longrightarrow$\hspace*{-.1cm} A\hspace*{-.1cm} =\hspace*{-.1cm} B
}
\end{flushleft}
\vspace*{-0.05cm}
 
\par We also extend the JNF library by formalizing additional matrix concepts and identities that support both the modeling and verification of our proposed formalization. For instance, we formalize diagonal and symmetric matrices as follows:

\begin{flushleft}
{\fontfamily{cmtt}\selectfont \small {\bf{definition}}\hspace*{-.1cm} \ka{diag\textunderscore matrix}\hspace*{-.1cm} ::\hspace*{-.1cm} $'$a\hspace*{-.1cm} ::\hspace*{-.1cm} zero\hspace*{-.1cm} mat\hspace*{-.1cm} $\Rightarrow$\hspace*{-.1cm} $'$a mat\hspace*{-.1cm} {\bf{where}} \\ \hspace*{.5cm} \ka{diag\textunderscore matrix}\hspace*{-.1cm} A\hspace*{-.1cm} $\equiv$\hspace*{-.1cm} (\ka{let} m = dim\_row\hspace*{-.1cm} A\hspace*{-.1cm} \\ \hspace*{4cm} \ka{in} mat\hspace*{-.1cm} m\hspace*{-.1cm} m\hspace*{-.1cm} ($\lambda$(i,j).\hspace*{-.1cm} if\hspace*{-.1cm} i\hspace*{-.1cm} =\hspace*{-.1cm} j\hspace*{-.1cm} then A\hspace*{-.1cm} $\$\$$\hspace*{-.1cm} (i,j) \hspace*{-.1cm}else\hspace*{-.1cm} 0))}
\end{flushleft}

It is worth noting that in the JNF library, a diagonal matrix (\texttt{diagonal\_mat}) is formalized as a predicate, which we use to verify necessary matrix identities. In contrast, we adopt a functional representation here, which is required for the formalization of degree matrices, as discussed in subsequent sections. Similarly, we formally define \texttt{is\_symmetric}, a square matrix whose transpose equals itself, as follows:

\begin{flushleft}
{\fontfamily{cmtt}\selectfont \small {\bf{definition}}\hspace*{-.1cm} \ka{is\_symmetric}\hspace*{-.1cm} ::\hspace*{-.1cm} $'$a\hspace*{-.1cm} mat\hspace*{-.1cm} $\Rightarrow$\hspace*{-.1cm} bool\hspace*{-.1cm} {\bf{where}} \\ \hspace*{.5cm} \ka{is\_symmetric}\hspace*{-.1cm} A\hspace*{-.1cm} $\equiv$\hspace*{-.1cm} (A\hspace*{-.1cm} =\hspace*{-.1cm} transpose\_mat\hspace*{-.1cm} A)\hspace*{-.1cm} $\land$\hspace*{-.1cm} square\_mat\hspace*{-.1cm} A}
\end{flushleft}

More details about the JNF matrix library and its extended version can be found in Isabelle's AFP~\cite{b21,JNFAFP}.

\section{Formalization of Network Systems}
\label{sec4graph}
\vspace*{-.1cm}
In this section, we introduce network systems capturing various graph representations using \textit{locales} in Isabelle/HOL. In particular, we provide the formalization of several fundamental graph types based on their topological characteristics, which is further used to formalize their matrix representation. Moreover, we explore the connection between our system with the most relevant existing graph theory in Isabelle/HOL.

A network system captured by a directed graph is  represented as an ordered pair $\mathcal{G}$=($\mathcal{N}$, $\mathcal{E}$), where $\mathcal{N}$ is the set of nodes and $\mathcal{E}$ is the set of directed edges. Each directed edge, denoted by $(n_{i}, n_{j})$, represents a pair of nodes, where $n_{i}$ is the head (starting node) and $n_{j}$ is the tail (ending node) of the edge. A weighted directed graph $\mathcal{G}_{\omega}$=($\mathcal{N}$, $\mathcal{E}$, $\omega$) is an extension of a directed graph by adding a weight function denoted by $\omega$. Graphs exhibit several key structural characteristics, such as self-loopness, symmetry and weight-balanced. For instance, an edge that connects a node to itself, denoted by $(n_{i}, n_{i})$, is known as a self-loop. Likewise, a graph is said to be weight-balanced if the weighted out- and in-degrees of every node are the same. These properties facilitate the understanding of the model’s structural behavior, which can be further analyzed algebraically using matrix representations.

To formalize network systems, 
we employ \textit{locales} in Isabelle/HOL, as introduced in Section~\ref{SEC:preliminaries}, which enables a modular transition from generic to specialized structures. We begin by defining generic network systems with minimal constraints (e.g., allowing self-loops, symmetry, etc.) and subsequently extend the locale to represent various specialized network types. This approach enables each specialized system to inherit definitions and theorems established in the more general locales. While our formalization leverages existing locale-based approaches in Isabelle/HOL, including \cite{noschinski2015graph}, \cite{b22} and \cite{kosaian2023first}, it differs in structure and objectives. For example, in~\cite{noschinski2015graph}, directed graphs are formalized using a record + locale structure, without incorporating weighted directed graphs. In~\cite{kovacs2020formalizing}, weighted undirected graphs are formalized via a locale restricted to loopless and symmetric structures, with node types constrained to linear orders and edge weights limited to $[0,\frac{q}{2}]\subset \mathbb{Z}$, where $q$ is the number of edges. In contrast, our network system formalization generalizes the structure, allowing self-loops and supporting both asymmetric and symmetric relationships, with generic node and edge types and positive real edge weights. Similarly, in~\cite{b22}, designs (generalizations of undirected graphs) are formalized using a core locale incidence system, with points as sets and blocks as multisets. The ordered incidence system locale further represents blocks as lists, facilitating the verification of incidence matrix properties in~\cite{b21}. Inspired by this approach, we use lists to model network systems, as their ordered elements simplify the verification of network topology matrices derived from the graph representation and their classical properties in Isabelle/HOL.

We start formalization of network systems captured by directed graphs with the locale \texttt{netw\_sys} which is composed of the parameters of the graph (the list of nodes ($\mathcal{N}$s) and edges ($\mathcal{E}$s)) and their relationships presented as well-formed assumptions. This locale ensures that all edges are pairs of nodes and the network has a distinct node list. The former is provided in the assumption \texttt{network\_wf} of the locale, while the latter is formalized in the assumption \texttt{distincts}.
\begin{flushleft}
{\fontfamily{cmtt}\selectfont \small {\bf{locale}} \ka{netw\textunderscore sys} = \\
\hspace{0.08in}{\bf{fixes}} nodes\textunderscore list\hspace*{-.1cm} ::\hspace*{-.1cm} $'$a\hspace*{-.1cm} nodes\hspace*{-.1cm} ($\mathcal{N}$s) 
{\bf{and}} edges\textunderscore list\hspace*{-.1cm} ::\hspace*{-.1cm}  $'$a edges\hspace*{-.1cm} ($\mathcal{E}$s) \\
\hspace{0.08in}{\bf{assumes}} network\textunderscore wf:\hspace*{-.1cm} $\bigwedge$e.\hspace*{-.1cm} e $\in$\hspace*{-.1cm} set\hspace*{-.1cm} $\mathcal{E}$s\hspace*{-.1cm} $\Longrightarrow$\hspace*{-.1cm} fst\hspace*{-.1cm} e\hspace*{-.1cm} $\in$\hspace*{-.1cm} set\hspace*{-.1cm} $\mathcal{N}$s\hspace*{-.1cm}
$\land$\hspace*{-.1cm} snd\hspace*{-.1cm} e\hspace*{-.1cm} $\in$\hspace*{-.1cm} set\hspace*{-.1cm} $\mathcal{N}$s \\
 \hspace{0.08in}{\bf{and}} distincts:\hspace*{-.1cm} distinct $\mathcal{N}$s
}
\end{flushleft}
\noindent where the function {\fontfamily{lmtt}\selectfont set} accepts a list and returns a set. Similarly, the function {\fontfamily{lmtt}\selectfont distinct} takes a list and ensures that elements of the list are disjoint. Furthermore, the functions \texttt{fst} and \texttt{snd} extract the first and second elements of a pair, respectively. For a better readability and usability in our proposed formalization of network topology matrices, we define the following type synonyms for nodes and edges:

\begin{flushleft}
{\fontfamily{cmtt}\selectfont \small {\bf{type\_synonym}} $'$a node = $'$a \hspace*{1cm}
{\bf{type\_synonym}} $'$a nodes = $'$a list \\
{\bf{type\_synonym}} $'$a edge = $'$a\hspace*{-.1cm} $\times$\hspace*{-.1cm} $'$a \hspace*{.1cm}
{\bf{type\_synonym}} $'$a edges = ($'$a\hspace*{-.1cm} $\times$\hspace*{-.1cm} $'$a)\hspace*{-.1cm} list} 
\end{flushleft}
Here, the type variable \texttt{$'$a} allows for a generic representation of graphs. It can be instantiated with any type, such as \texttt{nat} or \texttt{char list}, to label nodes and edges of the graph. We also provide some useful abbreviations for the set of nodes and edges and their cardinality.
\begin{flushleft}
{\fontfamily{cmtt}\selectfont \small {\bf{abbreviation}} m $\equiv$ length $\mathcal{N}$s \hspace*{1cm}
{\bf{abbreviation}} $\mathcal{N}$ $\equiv$ set $\mathcal{N}$s\\
{\bf{abbreviation}} n $\equiv$ length $\mathcal{E}$s  \hspace*{1.12cm}
{\bf{abbreviation}} $\mathcal{E}$ $\equiv$ set $\mathcal{E}$s} 
\end{flushleft}
To obtain a valid network system, we assume that lists of nodes and edges of the network should be nonempty. We formalize this condition in Isabelle/HOL as follows:
\vspace*{-.1cm}
\begin{flushleft}
{\fontfamily{cmtt}\selectfont \small {\bf{locale}} \ka{nonempty\_netw\_sys} = netw\_sys + \\
\hspace{0.05in}{\bf{assumes}} edges\textunderscore nempty:\hspace*{-.1cm} $\mathcal{E}$s $\neq$ [ ]} 
\end{flushleft}

Building upon the locale \texttt{nonempty\_netw\_sys}, the locale \texttt{sym\_netw\_sys} is formalized to ensure the existence of a reverse connection for every edge within the system.

\begin{flushleft}
{\fontfamily{cmtt}\selectfont \small {\bf{locale}} \ka{sym\_netw\_sys} = nonempty\_netw\_sys + \\
\hspace*{0.1cm}{\bf{assumes}} symmetric:\hspace*{-.1cm} $\forall$e $\in$ $\mathcal{E}$.\hspace*{-.1cm} (snd e, fst e) $\in$ $\mathcal{E}$}
\end{flushleft}

Similarly, a network system with no multi-edges is formalized as the following locale:
 
\begin{flushleft}
{\fontfamily{cmtt}\selectfont \small {\bf{locale}} \ka{nomulti\textunderscore netw\textunderscore sys} = noempty\_netw\textunderscore sys+ \\
\hspace*{0.05in}{\bf{assumes}} no\_multi\_edg:\hspace*{-.1cm} distinct $\mathcal{E}$s} 
\end{flushleft}

Here, we add the assumption that all edges in the network system are distinct. A simple network is then defined as a no-self loop network without multi-edges, which is formalized in the following locale as: 

\begin{flushleft}
{\fontfamily{cmtt}\selectfont \small {\bf{locale}} \ka{simple\_netw\textunderscore sys} = nomulti\_netw\textunderscore sys + \\
\hspace{0.05in}{\bf{assumes}} no\_self\_loop:\hspace*{-.1cm} $\bigwedge$\hspace*{-.1cm} e.\hspace*{-.1cm} e $\in$ $\mathcal{E}$ $\Longrightarrow$ fst e $\neq$ snd e} 
\end{flushleft}

This is achieved by incorporating the \texttt{no\_self\_loop} assumption, which asserts that the head and tail of each edge in the graph are distinct.

\par Now, we formally develop a weighted directed graph structure, defined by a locale \texttt{wdg\_sys}, as follows: 

\begin{flushleft}
{\fontfamily{cmtt}\selectfont \small {\bf{locale}} \ka{wdg\textunderscore sys} = nomulti\_netw\textunderscore sys + \\
\hspace{0.05in}{\bf{fixes}} weight\hspace*{-.1cm} ::\hspace*{-.1cm} $'$a weig\hspace*{-.1cm} ("wei")\\
\hspace*{0.05in}{\bf{assumes}} positive:\hspace*{-.1cm} $\forall$e\hspace*{-.1cm} $\in$\hspace*{-.1cm} $\mathcal{E}$.\hspace*{-.1cm} wei\hspace*{-.1cm} e $>$ 0 
\\
\hspace*{0.05in}{\bf{and}} zero:\hspace*{-.1cm} $\forall$\hspace*{-.1cm} x\hspace*{-.1cm} y.\hspace*{-.1cm} (x,y)\hspace*{-.1cm} $\notin$\hspace*{-.1cm} $\mathcal{E}$ $\longleftrightarrow$ wei\hspace*{-.1cm} (x,y)\hspace*{-.1cm} =\hspace*{-.1cm} 0} 
\end{flushleft}

Here, the parameter \texttt{$'$a weig} is defined as a new type synonym for the weight function that takes each edge and returns its corresponding real value. The assumption \texttt{positive} asserts the condition that all weights are positive. We also introduce the assumption \texttt{zero}, which ensures that the weights of non-existent edges are assigned a value of zero. It is important to note that \texttt{wdg\_sys} locale allows self-loops in the graph, providing a more general formulation. Moreover, we can also define a self-loop-free variant, which will be used in Section~\ref{sec6incirel}.

A weighted directed graph in which all edges have unit weight, commonly referred to as an unweighted directed graph or simply a directed graph, is formalized as follows:
\begin{flushleft}
{\fontfamily{cmtt}\selectfont \small {\bf{locale}} \ka{binary\_wdg\_sys} = wdg\_sys + \\
\hspace{0.05in}{\bf{assumes}} wei\_1:\hspace*{-.1cm} $\forall$e\hspace*{-.1cm} $\in$\hspace*{-.1cm} $\mathcal{E}$.\hspace*{-.1cm} wei\hspace*{-.1cm} e\hspace*{-.1cm} =\hspace*{-.1cm} 1}
\end{flushleft}

Next, we formalize a weighted directed graph with partitioned nodes , which enables the definition and analysis of subsystems over subsets of nodes. The utility of this formalization is further demonstrated in a relevant application in Section~\ref{sec7app}. This is specified in the following locale:
\vspace*{-.1cm}
\begin{flushleft}
{\fontfamily{cmtt}\selectfont \small {\bf{locale}} \ka{partitioned\_wdg\_sys} = wdg\_sys + \\
\hspace{0.05in}{\bf{fixes}} N1 N2\\
\hspace{0.05in}{\bf{assumes}} subnodes:\hspace*{-.1cm} $\mathcal{N}$s\hspace*{-.1cm} =\hspace*{-.1cm} N1\hspace*{-.1cm} $@$\hspace*{-.1cm} N2\hspace*{-.1cm} {\bf{and}}\hspace*{-.1cm} sub\_ge2:\hspace*{-.1cm} length\hspace*{-.1cm} N1\hspace*{-.1cm} $\ge$\hspace*{-.1cm} 2\hspace*{-.1cm}  {\bf{and}}\hspace*{-.1cm} n2\_ne:\hspace*{-.1cm} N2 $\neq$ [] }
\end{flushleft}

Similarly, the symmetric weighted directed graph, also refers to weighted undirected graph, is established through the following locale in Isabelle/HOL:

\begin{flushleft}
{\fontfamily{cmtt}\selectfont \small {\bf{locale}} \ka{sym\_wdg\_sys} = wdg\_sys + sym\_netw\_sys + \\
\hspace{0.05in}{\bf{assumes}} sym\_weight:\hspace*{-.1cm} $\forall$e\hspace*{-.1cm} $\in$\hspace*{-.1cm} $\mathcal{E}$.\hspace*{-.1cm} wei\hspace*{-.1cm} e\hspace*{-.1cm} =\hspace*{-.1cm} wei\hspace*{-.1cm} (snd\hspace*{-.1cm} e,\hspace*{-.1cm} fst\hspace*{-.1cm} e)}
\end{flushleft}

\begin{figure}[ht]
    \centering
    \includegraphics[width=.8\linewidth]{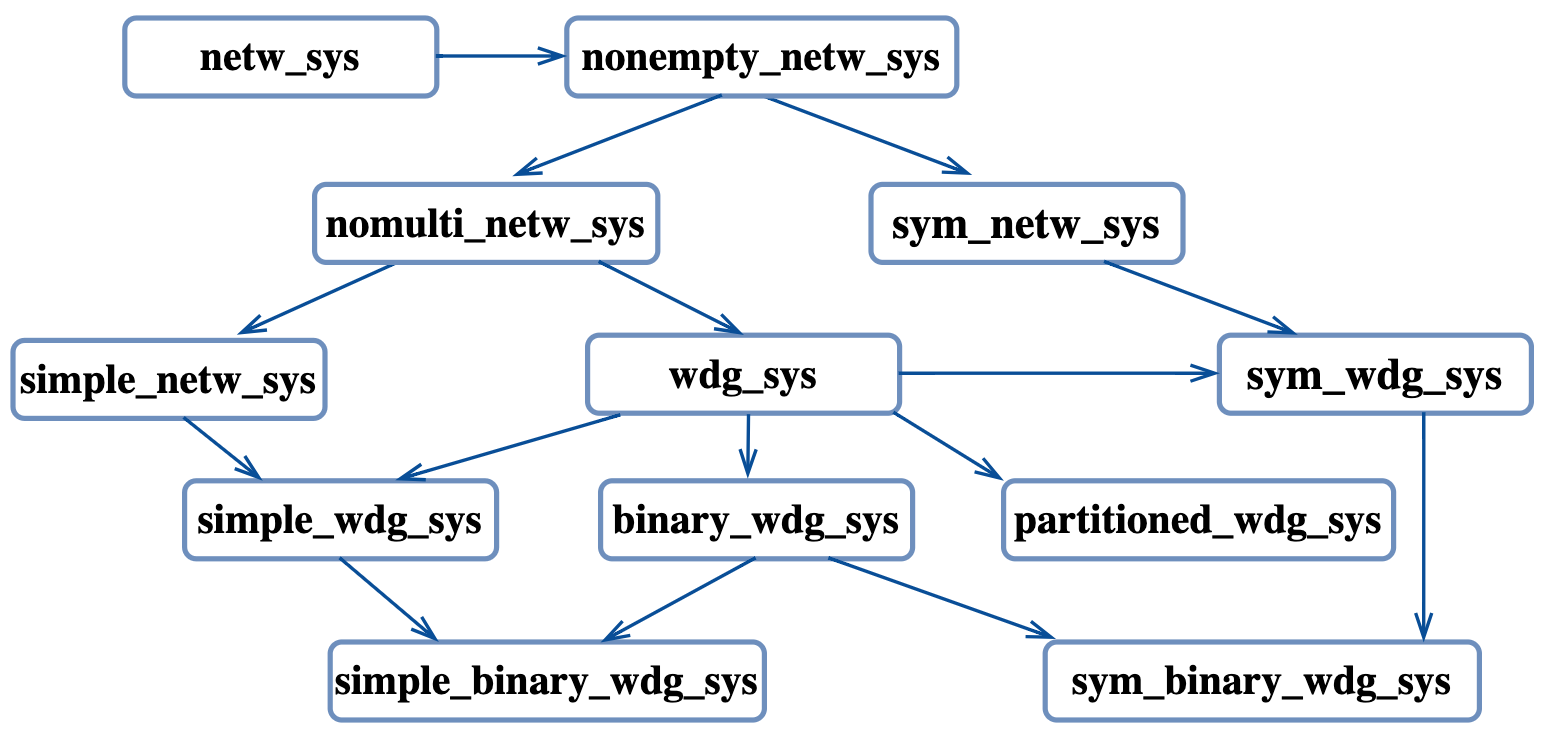} 
    \caption{Locales Development of Network Systems}
  \label{fig:rellocal} 
\end{figure}

Figure~\ref{fig:rellocal} illustrates the overall locale inheritance development, where locales are represented as rectangular boxes and their direct relationships with blue arrows. For example, the locale \texttt{simple\_wdg\_sys} inherits all definitions, assumptions and lemmas from the locales \texttt{wdg\_sys} and \texttt{simple\_netw\_sys}, allowing us to reuse verified lemmas from these locales. Similarly, the locale \texttt{wdg\_sys} is also extended to  
the locales \texttt{binary\_wdg\_sys} and \texttt{sym\_wdg\_sys}, representing the weighted directed and weighted undirected graphs, respectively. Furthermore, it should be noted that we modified and extended the network systems locales building upon our previous work~\cite{aksoy2025faecnttp}. In this paper, we introduce new graph concepts, e.g., \textit{weighted, symmetric, binary and partitioned} as extensions of the earlier framework, which included only the locales \texttt{netw\_sys}, \texttt{nonempty\_netw\_\\sys}, \texttt{nomulti\_netw\_sys}, and \texttt{noself\_netw\_sys}. 
\vspace*{-.1cm}
\vspace{-.5cm}  

\section{Formalization of Adjacency, Degree and Laplacian \break  Matrices}
\label{sec5adl}

In this section, we formalize several network topology matrices, including the adjacency, degree and Laplacian  matrices for \textit{weighted} directed graphs. These matrices can also represent other types of graphs, such as undirected and or unweighted graphs, and can be defined over real or complex numbers depending on the application. To achieve a more generic formalization and enhance applicability, we define the matrices over a field, independently of any locale, and then demonstrate their equivalent definitions within the weighted directed graph locale.
\vspace*{-.2cm}
\subsection{Adjacency Matrices}
\label{sec5adj}
Adjacency matrices are used to represent the relationships between nodes in a graph and to determine the existence of edges. Different variations of adjacency matrices exist in the literature, depending on the type of graph. For instance, an adjacency matrix of a network captured by a weighted directed graph with $m$ nodes and no multi-edges is defined as follows:  

\begin{definition}
\label{DEF:adj} 
{\textup{\textit{Adjacency Matrix of a Weighted Directed Graph}}~\cite{bullo2018lectures}} \\
\normalfont Consider a network with a set of nodes $\mathcal{N}=\{x_{1}, x_{2}, \dots, x_{m}\}$ and a set of edges $\mathcal{E} \subseteq \mathcal{N}\times \mathcal{N}$, the corresponding adjacency matrix $\mathcal{A} = [a_{ij}]$ is defined by an $m \times m$ matrix:
\begin{equation}
 a_{ij} = \left\{ \begin{array}{lll}
 \omega (x_{i}, x_{j}), & \textrm{if there is an edge directed from node $x_{i}$ to node $x_{j}$} \\ 
 0, & \textrm{otherwise} \\
\end{array} \right.   
\end{equation}
\end{definition}
\noindent for $i, j \in \{1,2,\dots,m\}$. Here, $\omega : \mathcal{E} \rightarrow \mathbb{R} $ is a weight function accepting an edge as an input and returns its corresponding weight as an output. 

\par
\begin{figure}[htbp]
  \subcaptionbox*{(a) IEEE 5-Bus System }[.33\linewidth]{%
    \includegraphics[width=1\linewidth]{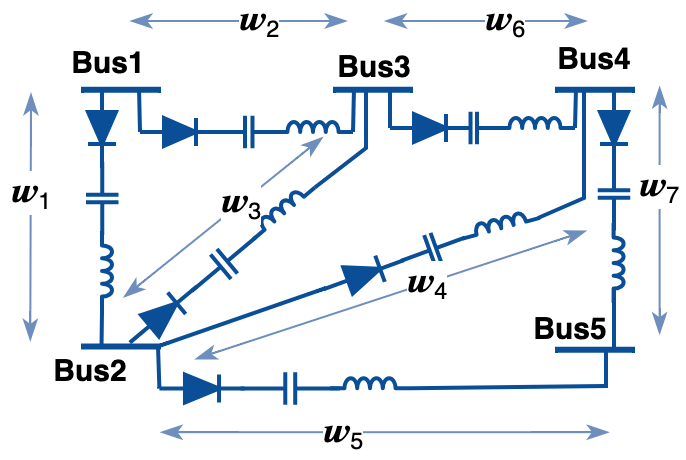}%
     \label{Fig: ieee5}
  }%
  \hfill 
  \subcaptionbox*{(b) Weighted Directed Graph}[.34\linewidth]{%
    \includegraphics[width=1\linewidth]{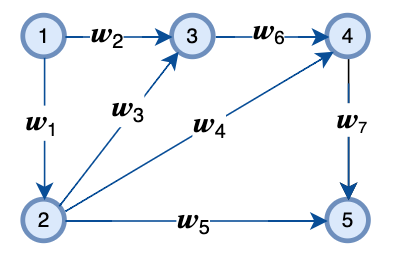}%
     \label{Fig: weidg}
  }%
  \hfill
  \subcaptionbox*{(c) Adjacency Matrix}[.33\linewidth]{%
    \includegraphics[width=1.1\linewidth]{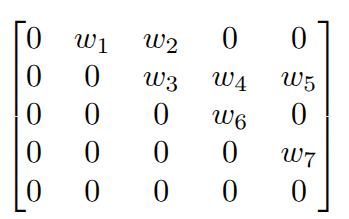}%
     \label{Fig: ieeeadj}
  }
  
  \caption{IEEE 5-Bus System with its Weighted Directed Graph Representation and Adjacency Matrix}
  \label{Fig:weidgad}
\end{figure}

 When the weight function is equal to one, i.e., $\omega (x_{i}, x_{j}) = 1$, if there is an edge directed from node $x_{i}$ to node $x_{j}$, the above-presented weighted adjacency matrix is transformed to a binary adjacency matrix. Figure~\ref{Fig:weidgad} depicts the concept of representing a power system as a graph and its adjacency matrix using the IEEE 5-Bus test system, a standard benchmark widely used in power flow analysis.  The graph consists of nodes $\{1,2,\dots,5\}$ and edges with weights $\{w_{1},w_{2},\dots,w_{7}\}$, while the non-zero entries of the adjacency matrix indicate the presence of edges and their associated weights (e.g., $w_{1}$ represents the weight of the directed edge from node 1 to node 2).

\par We first formalize the adjacency matrix over a field in Isabelle/HOL as follows: 
\begin{flushleft}
{\fontfamily{cmtt}\selectfont \small {\bf{definition}} 
\ka{adjacency\_mat}\hspace*{-.1cm} ::\hspace*{-.1cm} $'$a\hspace*{-.1cm} nodes\hspace*{-.1cm} $\Rightarrow$\hspace*{-.1cm} $'$a\hspace*{-.1cm} edges\\ \hspace*{4.5cm}  $\Rightarrow$ ($'$a\hspace*{-.1cm} edge\hspace*{-.1cm} $\Rightarrow$\hspace*{-.1cm} $'$b)\hspace*{-.1cm} $\Rightarrow$\hspace*{-.1cm} $'$b\hspace*{-.1cm} ::\hspace*{-.1cm} field\hspace*{-.1cm} mat
\\ \hspace*{.1cm}
{\bf{where}} \ka{adjacency\_mat}\hspace*{-.1cm} Ns\hspace*{-.1cm} Es\hspace*{-.1cm} wei\hspace*{-.1cm} $\equiv$\hspace*{-.1cm} mat\hspace*{-.1cm} (length\hspace*{-.1cm} Ns)\hspace*{-.1cm} (length\hspace*{-.1cm} Ns)
\\ \hspace*{2.5cm} ($\lambda$(i,j).\hspace*{-.1cm} if\hspace*{-.1cm} (Ns!i,\hspace*{-.1cm} Ns!j)\hspace*{-.1cm} $\in$\hspace*{-.1cm} set\hspace*{-.1cm} Es\hspace*{-.1cm} then\hspace*{-.1cm} wei\hspace*{-.1cm} (Ns!i,\hspace*{-.1cm} Ns!j)\hspace*{-.1cm} else\hspace*{-.1cm} 0)}
\end{flushleft}
\noindent Here, \texttt{wei} represents the weight function with type  \texttt{$'$a\hspace*{-.1cm} edge\hspace*{-.1cm} $\Rightarrow$\hspace*{-.1cm} $'$b}. To ensure generality and reusability, the matrix type \texttt{$'$b mat} is an instance of the \texttt{field} type class, allowing instantiation with  common matrix types such as real numbers ($\mathbb{R}$) and complex numbers ($\mathbb{C}$). For instance, within the locale \texttt{wdg\_sys}, we use $\mathpzc{A}$ to abbreviate the adjacency matrix over real numbers corresponding to a weighted directed graph (Definition~\ref{DEF:adj}) as follows: 
\vspace*{-.1cm} 
\begin{flushleft}
{\fontfamily{cmtt}\selectfont \small {\bf{abbreviation}} 
$\mathpzc{A}$\hspace*{-.05cm} ::\hspace*{-.05cm} real\hspace*{-.05cm} mat\hspace*{-.05cm} {\bf{where}}\hspace*{-.05cm} $\mathpzc{A}$\hspace*{-.05cm} $\equiv$\hspace*{-.05cm} adjacency\_mat\hspace*{-.05cm} $\mathcal{N}$s\hspace*{-.05cm} $\mathcal{E}$s\hspace*{-.05cm} wei}
\end{flushleft}
\vspace*{-.05cm} 
We can then quickly verify an explicit representation of the adjacency matrix within the locale \texttt{wdg\_sys} as follows:
\begin{flushleft}
{\fontfamily{cmtt}\selectfont \small {\bf{lemma}} 
\ka{adj\_alt}\hspace*{-.1cm} :\hspace*{-.1cm} $\mathpzc{A}$\hspace*{-.05cm} $\equiv$\hspace*{-.1cm} mat\hspace*{-.1cm} m\hspace*{-.1cm} m\hspace*{-.1cm} ($\lambda$(i,j).\hspace*{-.1cm}  wei\hspace*{-.1cm} ($\mathcal{N}$s!i,\hspace*{-.1cm} $\mathcal{N}$s!j))}
\end{flushleft}
 \noindent This exemplifies how locale assumptions enable us to simplify the definitions within the locale contexts, which in turn leads to easier proofs. We can also formally verify a binary adjacency matrix that represents an unweighted graph as a 0-1 matrix within the locale \texttt{binary\_wdg\_sys}, as follows:
 
\begin{flushleft}  
{\fontfamily{cmtt}\selectfont \small {\bf{lemma}} \ka{binary\textunderscore is\_wei\_1}: \\ \hspace*{1cm} $\mathpzc{A}$\hspace*{-.1cm} =\hspace*{-.1cm} mat\hspace*{-.1cm} m\hspace*{-.1cm} m\hspace*{-.1cm} ($\lambda$(i,j).\hspace*{-.1cm} if\hspace*{-.1cm} ($\mathcal{N}$s!i,\hspace*{-.1cm} $\mathcal{N}$s!j)\hspace*{-.1cm} $\in$\hspace*{-.1cm} $\mathcal{E}$\hspace*{-.05cm} then\hspace*{-.05cm} 1\hspace*{-.05cm} else\hspace*{-.05cm} 0)}
\end{flushleft}

Moreover, we verify the symmetry property of the adjacency matrix corresponding to an undirected weighted graph within the locale \texttt{sym\_wdg\_sys}. 
\vspace*{-.01cm}
\begin{flushleft}  
 {\fontfamily{cmtt}\selectfont \small {\bf{lemma}} \ka{adjmat\_sym}:\hspace*{-.1cm} is\_symmetric $\mathpzc{A}$}
\end{flushleft}

The adjacency matrix also provides a basis to obtain weighted out- and in-degrees. Mathematically, the weighted out-degree of a node is defined as the sum of weights of edges that are outgoing from that node, while weighted in-degree of a node is the sum of weights of edges that are incoming to that node. An edge can be represented by a pair of nodes (\texttt{$\mathcal{N}$s$!$i,$\mathcal{N}$s$!$j}) or through an element of the edge list (\texttt{$\mathcal{E}$s$!$k}). We refer to the former as explicit representation, while the latter is an implicit representation. For instance, the out-degree notion, based on the explicit representation of edges, is formalized in Isabelle/HOL as follows: 

\begin{flushleft}  
 {\fontfamily{cmtt}\selectfont \small {\bf{definition}} \ka{wei\_outdegree}\hspace*{-.1cm} ::\hspace*{-.1cm} $'$a\hspace*{-.1cm} $\Rightarrow$\hspace*{-.1cm} real {\bf{where}} \\
\hspace*{1cm}\ka{wei\_outdegree}\hspace*{-.1cm} u\hspace*{-.1cm} $\equiv$\hspace*{-.1cm} $\sum$\hspace*{-.1cm} j\hspace*{-.1cm} $\in$\hspace*{-.1cm} $\{$j\hspace*{-.1cm} $\in$\hspace*{-.1cm} $\{$0..$<$\hspace*{-.1cm} m$\}$.\hspace*{-.1cm} (u,\hspace*{-.1cm} $\mathcal{N}$s!j)\hspace*{-.1cm} $\in$\hspace*{-.1cm} $\mathcal{E}$$\}$.\hspace*{-.1cm} wei\hspace*{-.1cm} (u,\hspace*{-.1cm} $\mathcal{N}$s!j)
 }
\end{flushleft}

\noindent Here, \texttt{wei\_outdegree} accepts a node \texttt{u} as an input and returns the sum of weights of edges, where \texttt{u} represents the first element of each of the edges, captured as pair of nodes. 

Similarly, \texttt{wei\_outdegree} takes a node \texttt{v} as an input and returns the sum of weights of edges that are incoming to the node \texttt{v}. Alternatively, we also formalize the weighted degrees (out- and in-degrees) using the implicit form of the edges, which can be found  in the "\textit{Network\_Systems}" theory~\cite{b25}. This concept can make the reasoning easier when dealing with the models constructed through the node-edge relationships.

We now verify two important lemmas, where the weighted in- and out-degrees are obtained using the adjacency matrix in Isabelle/HOL.

\begin{flushleft}
{\fontfamily{cmtt}\selectfont \small {\bf{lemma}} \ka{wei\_out\_adj}:\hspace*{-.1cm} i$<$\hspace*{-.1cm} m $\Longrightarrow$\\ \hspace*{1cm} wei\_outdegree\hspace*{-.1cm} ($\mathcal{N}$s!i)\hspace*{-.1cm} = $\sum$\hspace*{-.1cm} j\hspace*{-.1cm} $\in$\hspace*{-.1cm} $\{$0..$<$\hspace*{-.1cm} m$\}$.\hspace*{-.1cm}  $\mathpzc{A}$\hspace*{-.1cm} $\$\$$\hspace*{-.1cm} (i,j)}

\end{flushleft}

\noindent The verification of the above lemma is based on the definitions of \texttt{wei\_outdegree} and \texttt{adjacency\_mat} alongside the matrix indexing and summation reasoning.

\begin{flushleft}
{\fontfamily{cmtt}\selectfont \small {\bf{lemma}} \ka{wei\_in\_adj}:\hspace*{-.1cm} i$<$\hspace*{-.1cm} m $\Longrightarrow$\\ \hspace*{1cm} wei\_indegree\hspace*{-.1cm} ($\mathcal{N}$s!i)\hspace*{-.1cm} = $\sum$\hspace*{-.1cm}  j\hspace*{-.1cm} $\in$\hspace*{-.1cm} $\{$0..$<$\hspace*{-.1cm} m$\}$.\hspace*{-.1cm} $\mathpzc{A}^{T}$\hspace*{-.1cm} $\$\$$\hspace*{-.1cm} (i,j)}

\end{flushleft}

\noindent The proof process of the above lemma is very similar to that of the previous lemma. 

\subsection{Degree Matrices}

Degree matrices are diagonal matrices that characterize node connectivity in a weighted directed graph. They include the out-degree and in-degree matrices, whose diagonal elements represent the corresponding degree of each node. For a weighted directed graph with 
$m$ nodes, these matrices are defined in terms of the adjacency matrix as follows:

\begin{definition}{\textit{Degree Matrices of a Weighted Directed Graph~\cite{bullo2018lectures}}}\\
\label{DEF:dindout}
\normalfont The out-degree $\mathcal{D}_{out}$ and in-degree $\mathcal{D}_{in}$ matrices are defined as $m\times m$ diagonal matrices:
\begin{align*}
     \mathcal{D}_{out} &= diag (\mathcal{A} * \mathbb{J}_{m})\\
       \mathcal{D}_{in} &= diag (\mathcal{A}^T * \mathbb{J}_{m}) 
\end{align*}
\end{definition}
\noindent Here, the function $diag: \mathbb{R} \rightarrow \mathbb{R}^{m\times m}$ maps the real-valued weights to the diagonal elements of an $m \times m$ matrix with non-diagonal elements set to zero. Additionally, $\mathbb{J}_{m}$ represents an $m\times m$ matrix with all elements equal to 1. 

\begin{figure}[htbp]
  \subcaptionbox*{\hspace*{-.6cm}(a) Out-degree Matrix}[.32\linewidth]{%
   \hspace*{-.31cm} \includegraphics[width=.91\linewidth]{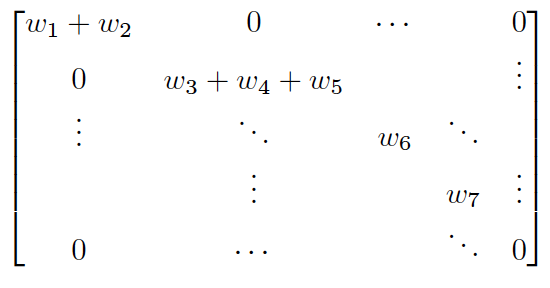}%
     \label{Fig: doutmat}
  }%
  \hfill
  \subcaptionbox*{\hspace*{-.7cm}(b) In-degree Matrix}[.32\linewidth]{%
    \hspace*{-.31cm}\includegraphics[width=.91\linewidth]{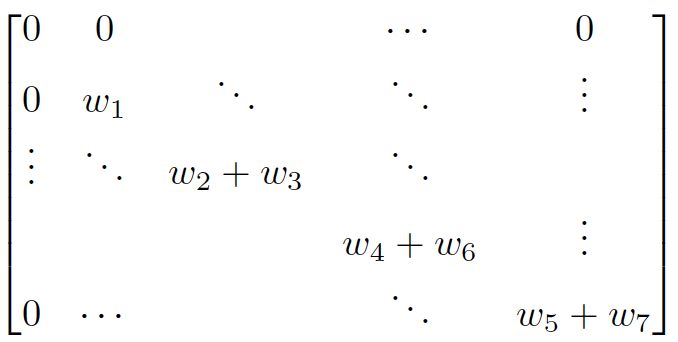}%
     \label{Fig: imat}
     }%
  \hfill
  \subcaptionbox*{\hspace*{-.6cm}(c) Laplacian Matrix}[.35\linewidth]{%
    \hspace*{-.31cm}\includegraphics[width=1\linewidth]{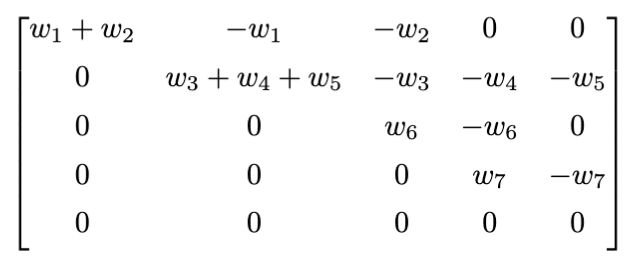}%
     \label{Fig: iimat}
  }
  {\caption{Degree and Laplacian Matrices of the Graph in Figure~\ref{Fig:weidgad}\textcolor{blue3}{(b)}}
  \label{Fig: degrees}}
  \vspace*{-.5cm}
\end{figure}

Figure~\ref{Fig: degrees} provides the in-degree and out-degree matrices of the weighted directed graph representing the IEEE $5$-Bus system, shown in Figure~\ref{Fig:weidgad}. These matrices are diagonal, with each diagonal element computed as the sum of the corresponding row or column of the  adjacency matrix. For example, node $1$ has two outgoing edges with weights, $w_{1}$ and $w_{2}$, so its corresponding entry in the out-degree matrix is the sum of these weights, which corresponds to the sum of the entries in first row of the adjacency matrix. Conversely, since node $1$ has no incoming weighted edges, its corresponding entry in the in-degree matrix is $0$, matching the sum of first column. From this matrices, it can also be observed that the system is neither weight-balanced nor symmetric. 

We now formalize the out-degree matrix for an arbitrary list of nodes and edges with weights using Isabelle/HOL as follows:
\vspace*{-.05cm} 
\begin{flushleft}
{\fontfamily{cmtt}\selectfont \small {\bf{definition}} 
\ka{out\_degree\_mat}\hspace*{-.1cm} {\bf{where}} \\ \hspace*{1cm} \ka{out\_degree\_mat}\hspace*{-.1cm}  Ns\hspace*{-.1cm} Es\hspace*{-.1cm} wei $\equiv$\hspace*{-.1cm} diag\_matrix\hspace*{-.1cm} (outhelper Ns Es wei)}
\end{flushleft}

\noindent where \texttt{outhelper} is defined as the product of the adjacency matrix and $\mathbb{J}_{m}$ matrix, and is formalized as follows:  
\vspace*{-.05cm} 
\begin{flushleft}
{\fontfamily{cmtt}\selectfont \small {\bf{definition}} 
\ka{outhelper}\hspace*{-.01cm} {\bf{where}} \\ \hspace*{1cm} \ka{outhelper}\hspace*{-.1cm} $\equiv$\hspace*{-.01cm} (adjacency\_mat\hspace*{-.1cm} Ns\hspace*{-.1cm} Es\hspace*{-.1cm} wei)\hspace*{-.1cm} *\hspace*{-.1cm} ($\mathbf{J_{m}}$\hspace*{-.1cm} (length\hspace*{-.1cm} Ns))}
\end{flushleft}
\vspace*{-.05cm} 

\noindent where $\mathbf{J}_{m}$ is an Isabelle/HOL function that accepts a dimension (e.g., \texttt{length Ns}) and constructs a square matrix of that dimension, where all elements are equal to one. Similarly, the in-degree matrix for an arbitrary number of nodes and edges is formalized as: 
\vspace*{-.05cm} 
\begin{flushleft}
{\fontfamily{cmtt}\selectfont \small {\bf{definition}} 
\ka{in\_degree\_mat}\hspace*{-.1cm} {\bf{where}} \\ \hspace*{1cm} \ka{in\_degree\_mat}\hspace*{-.1cm} Ns\hspace*{-.1cm} Es\hspace*{-.1cm} wei\hspace*{-.1cm} $\equiv$\hspace*{-.1cm} diag\_matrix\hspace*{-.1cm} (inhelper Ns Es wei)}
\end{flushleft}

\noindent where \texttt{inhelper} represents the product of the transpose of the adjacency matrix and $\mathbb{J}_{m}$ matrices. These definitions are globally accessible and used within the relevant network system locales. In the locale \texttt{wdg\_sys}, we abbreviate them as \texttt{$\mathpzc{D}$out} and \texttt{$\mathpzc{D}$in}, corresponding to Definition~\ref{DEF:dindout}. We can formally verify that the diagonal entries of the out-degree matrix equal the weighted out-degree of the corresponding nodes:
\vspace*{-.05cm} 
\begin{flushleft}
{\fontfamily{cmtt}\selectfont \small {\bf{lemma}} \ka{out\_deg\_mat\_wei\_out}: \\
\hspace*{.2cm} {\bf{assumes}} i\hspace*{-.1cm} $<$\hspace*{-.1cm} m {\bf{and}} i\hspace*{-.1cm} =\hspace*{-.1cm} j \\
\hspace*{.2cm} {\bf{shows}} $\mathpzc{D}$out $\$\$$ (i,j) = wei\_outdegree ($\mathcal{N}$s!i)}
\end{flushleft}

The proof of this lemma leverages the previously verified \texttt{wei\_out\_adj} property, which establishes the relationship between adjacency matrix and out-degree values,  along with degree index relationships.

Likewise, we verify that the diagonal elements of the in-degree matrix are equal to the weighted in-degree of the corresponding nodes, as stated in the following lemma:

\begin{flushleft}
{\fontfamily{cmtt}\selectfont \small {\bf{lemma}} \ka{in\_deg\_mat\_wei\_in}: \\
\hspace*{.2cm} {\bf{assumes}} i\hspace*{-.1cm} $<$\hspace*{-.1cm} m {\bf{and}} i\hspace*{-.1cm} =\hspace*{-.1cm} j \\
\hspace*{.2cm} {\bf{shows}} $\mathpzc{D}$in $\$\$$ (i,j) = wei\_indegree ($\mathcal{N}$s!i)}
\end{flushleft}
 
It is worth noting that the degree matrices are alternatively formalized using the implicit version of the degree concept~\cite{b25}, which is particularly useful when verifying relationships between incidence and degree matrices, as discussed in the next section.
\par Another useful topological characteristic for weighted graphs is weight-balance, where each node has equal weighted out- and in-degrees. That is, for a graph having weight-balanced nodes, its associated out-degree and in-degree matrices are identical. This result is verified as the following lemma:

\begin{flushleft}
{\fontfamily{cmtt}\selectfont \small {\bf{lemma}} \ka{wei\_balanced\_in\_out\_deg}: \\
\hspace*{.2cm} {\bf{assumes}} $\bigwedge$i.\hspace*{-.1cm} i\hspace*{-.1cm} $<$\hspace*{-.1cm} m\hspace*{-.1cm} $\Longrightarrow$\hspace*{-.1cm} weight\_balanced\hspace*{-.1cm} ($\mathcal{N}$s!i) \\
\hspace*{.2cm} {\bf{shows}} $\mathpzc{D}$out\hspace*{-.1cm} = $\mathpzc{D}$in}
\end{flushleft}

The proof of the above lemma proceeds by verifying three subgoals generated from the~\texttt{eq\_matI} introduction rule. After establishing the matrix size equality, element-wise equality is verified by considering the cases \texttt{i=j} and \texttt{i$\neq$j}, using the formerly stated lemmas \texttt{out\_deg\_mat\_wei\_out} and \texttt{in\_deg\_mat\_wei\_in}.

Additionally, for binary weighted directed graphs, each diagonal entry of the out-degree (or in-degree) matrix equals the sum of ones in the corresponding row (or column). More details about these formalizations are available in our proof script ~\cite{b25}.

\subsection{Laplacian Matrices}
\label{sec5lap}

For a network modeled as a weighted directed graph with $m$ nodes and $n$ edges, a Laplacian matrix is mathematically defined as follows:

\begin{definition}
\label{DEF:LAPMAT}Laplacian Matrix of a Weighted Directed Graph~\cite{bullo2018lectures} \\
\normalfont The Laplacian matrix $\mathcal{L}= [\ell_{ij}]$ of a weighted directed graph is defined by an $m \times m$ matrix, such that 
 
 \begin{equation}\label{LAP}
     \mathcal{L} = \mathcal{D}_{out} - \mathcal{A}
 \end{equation}

An $(i,j)$-th entry of the Laplacian matrix is given by:

\begin{equation}
    \ell_{ij} = 
    \begin{cases} 
        -a_{ij}, & \text{if } i \neq j \\ 
        \sum_{h=1, h \neq i}^{n} a_{ih}, & \text{if } i = j
    \end{cases}
\label{placeholder}
\end{equation}

\end{definition}

\noindent where $a_{ij}$ represents the $(i,j)$-th entry of the adjacency matrix. In this definition, if an edge is directed from node $i$ to node $j$, the corresponding off-diagonal entry is the negative of its weight, whereas the diagonal entries are equal to the sum of the weights of all edges outgoing from node $i$. For instance, Figure~\ref{Fig: degrees}\ka{(c)} illustrates the Laplacian matrix of the weighted directed graph corresponding to the IEEE 5-Bus system (depicted in Figure~\ref{Fig:weidgad}).  Using the formal definitions of the adjacency and the out-degree matrices, we formalize the Laplacian matrix for arbitrary lists of nodes and edges with a weight function in Isabelle/HOL as follows:
\vspace*{-.1cm}
\begin{flushleft}
{\fontfamily{cmtt}\selectfont \small {\bf{definition}} 
\ka{laplacian\_mat}\hspace*{-.1cm} {\bf{where}}\\ \hspace*{.2cm} \ka{laplacian\_mat}\hspace*{-.1cm} Ns\hspace*{-.1cm} Es\hspace*{-.1cm} wei\hspace*{-.1cm} $\equiv$\hspace*{-.1cm} out\_degree\_mat\hspace*{-.1cm} Ns\hspace*{-.1cm} Es\hspace*{-.1cm} wei\hspace*{-.1cm} -\hspace*{-.1cm} adjacency\_mat\hspace*{-.1cm} Ns\hspace*{-.1cm} Es\hspace*{-.1cm} wei
}
\end{flushleft}
\vspace*{-.1cm}
Within the locale \texttt{wdg\_sys}, the Laplacian matrix is denoted as $\mathpzc{L}$, and the following lemma corresponds to Equation~(\ref{LAP}) over reals:   
\vspace*{-.1cm}
\begin{flushleft}
{\fontfamily{cmtt}\selectfont \small {\bf{lemma}} 
\ka{L\_mat}: $\mathpzc{L}$ = $\mathpzc{D}$out  - $\mathpzc{A}$
}
\end{flushleft}
\vspace*{-.1cm}
Next, we verify the indexing properties, based on Equation~(\ref{placeholder}), as the following lemma in Isabelle/HOL:

\vspace*{-.2cm}
\begin{flushleft}
{\fontfamily{cmtt}\selectfont \small {\bf{lemma}} \ka{laplacian\_index}:\hspace*{-.1cm} \\
\hspace*{.7cm}{\bf{assumes}} i\hspace*{-.1cm} <\hspace*{-.1cm} m\hspace*{-.1cm} {\bf{and}}\hspace*{-.1cm} j\hspace*{-.1cm} <\hspace*{-.1cm} m\\
\hspace*{.7cm}{\bf{shows}}\hspace*{-.1cm} i$\neq$j\hspace*{-.1cm} $\Longrightarrow$\hspace*{-.1cm} $\mathpzc{L}$\hspace*{-.1cm} $\$\$$\hspace*{-.1cm} (i,j)=\hspace*{-.1cm} -\hspace*{-.1cm} $\mathpzc{A}$\hspace*{-.1cm} $\$\$$\hspace*{-.1cm} (i,j)\\
\hspace*{.9cm} {\bf{and}}\hspace*{-.1cm} i=j\hspace*{-.1cm} $\Longrightarrow$\hspace*{-.1cm} $\mathpzc{L}$\hspace*{-.1cm} $\$\$$\hspace*{-.1cm} (i,j)\hspace*{-.1cm} = $\sum$\hspace*{-.1cm} h\hspace*{-.1cm} $\in$\hspace*{-.1cm} $\{$0..<m$\}-\{$i$\}$.\hspace*{-.1cm} $\mathpzc{A}$\hspace*{-.1cm} $\$\$$\hspace*{-.1cm} (i,h)}
\end{flushleft}
\vspace*{-.2cm}
This indexing lemma verifies the entries of the Laplacian matrix by deriving its diagonal and off-diagonal elements from the adjacency matrix. Next, we present some key properties of the Laplacian matrix of a weighted directed graph, which are useful in modelling conservation laws (e.g., Kirchhoff's Current Law (KCL) and Ohm's Law) and analysis of physical phenomena, such as power flow and heat transfer.

We now introduce row and column related properties of the Laplacian matrix. Given a weighted directed graph with $m$ nodes and its corresponding Laplacian matrix $\mathcal{L}$, the sum of the rows of the Laplacian matrix is zero, which is mathematically expressed as~\cite{bullo2018lectures}: 

\begin{equation}
\label{lemmarowlap}
    \mathcal{L}\:\Vec{1}_{m} =\Vec{0}_{m}
\end{equation}

\noindent where $\Vec{1}_{m}$ represents an $m$-dimensional vector with all elements being one, and $\Vec{0}_{m}$ denotes an $m$-dimensional zero vector. This lemma also shows that $\Vec{1}_{m}$ is in the kernel of Laplacian matrix $\mathcal{L}$, which provides important insights for the study of the matrix spectrum\footnote{The set of all eigenvalues of a matrix is called a \textit{spectrum}.}. We now formally verify Equation~(\ref{lemmarowlap}) in Isabelle/HOL as the following lemma:

\begin{flushleft}
{\fontfamily{cmtt}\selectfont \small {\bf{lemma}} \ka{zero\_row\_sums}:\hspace*{-.1cm}\\ 
\hspace*{.01cm} {\bf{shows}}\hspace*{-.1cm} i\hspace*{-.1cm} <\hspace*{-.1cm} m\hspace*{-.1cm} $\Longrightarrow$ $\sum$\hspace*{-.1cm} j\hspace*{-.1cm} <\hspace*{-.1cm} m.\hspace*{-.1cm} $\mathpzc{L}$ $\$\$$ (i,j) = 0}
\end{flushleft}

\noindent where $\mathpzc{L}$ is the abbreviation for the Laplacian matrix. We verify the above lemma by decomposing the summation over the index set ($\{0,\dots, m-1\}$) based on diagonal and off-diagonal entries. This separation allows us to apply the Laplacian matrix indexing lemma to each case independently, and the result is obtained from summation properties. Similarly, the column-related property of the Laplacian matrix of a weighted directed graph with $m$ nodes is stated as the sum of the columns of the Laplacian matrix is zero if and only if the graph is weight-balanced. This can be mathematically expressed by the following equality: 

\begin{equation}
\label{lemmacol}
    \mathcal{D}_{in} = \mathcal{D}_{out} \Longleftrightarrow (\mathbb{J}_{m})^{T}\mathcal{L} = \mathbf{0}_{m}
\end{equation}

\noindent where $\mathbf{0}_{m}$ refers to an $m\times m$ zero-valued matrix. Here, the equality  $\mathcal{D}_{in} = \mathcal{D}_{out}$ represents that the given weighted directed graph is weight-balanced. We formally verify this property in Isabelle/HOL as follows:

\begin{flushleft}
{\fontfamily{cmtt}\selectfont \small {\bf{lemma}} \ka{zero\_col\_sums}:\hspace*{-.1cm}
\\ 
\hspace*{.2cm} 
{\bf{assumes}}\hspace*{-.1cm} i\hspace*{-.1cm} <\hspace*{-.1cm} m\hspace*{-.1cm} {\bf{and}}\hspace*{-.1cm} j\hspace*{-.1cm} <\hspace*{-.1cm} m 
\\
\hspace*{.2cm} {\bf{shows}}\hspace*{-.1cm} $\mathpzc{D}$in = $\mathpzc{D}$out  $\longleftrightarrow$ ($\mathbf{J}_{m}$\hspace*{-.1cm} m)$^{T}$\hspace*{-.1cm} *\hspace*{-.1cm} $\mathpzc{L}$\hspace*{-.1cm} =\hspace*{-.1cm} ($\mathbf{0}$$_{m}$\hspace*{-.1cm} m\hspace*{-.1cm} m)}
\end{flushleft}

The verification of the above lemma relies on algebraic properties related to the transpose, indexing and dimension of the associated matrices, along with the proof of the following equations.
\begin{equation}
\mathbb{J}_{m}^{T}\mathcal{L} = \mathcal{L}^{T}\mathbb{J}_{m} 
\label{helper1}
\end{equation}
\begin{equation} \mathcal{L}^{T}\mathbb{J}_{m} = \mathcal{D}_{out} - \mathcal{D}_{in}
 \label{helper3}
\end{equation}

 Table~\ref{tablemmacol} demonstrates some lemmas that are utilized to formally verify the sum of columns of the Laplacian matrix is equal to zero (Equation~(\ref{lemmacol})). For instance, Equation~(\ref{helper1}) is formalized in Isabelle/HOL as described in the first entry of the table, while the formalization of indexed version of Equation~(\ref{helper3}) corresponds to the second entry of the table. The third entry states that if a square matrix is diagonal, its product with a matrix having all elements equal to 1, yields the diagonal entries of that matrix.
\vspace*{-.3cm}
\begin{table} [htb]
\begin{center}
\resizebox{\textwidth}{!}{
\begin{tabular}{|l|l|} 
\hline \thead{\\ \makecell{\fontsize{18pt}{10pt}\selectfont \textbf{Mathematical Form}}\\} & \thead{\\ \fontsize{18pt}{10pt}\selectfont \textbf{Isabelle/HOL Formalization}\\} \\ 
\hline \hline
\makecell{\fontsize{18pt}{10pt}\selectfont$
  ($$\mathbb{J}_{m})^TA = \mathbf{0}_{m}
  \Longleftrightarrow
  A^T(\mathbb{J}_{m}) = \mathbf{0}_{m}$$
$} &
  \makecell[l]{\\ \texttt{\fontsize{17pt}{10pt}\selectfont {\bf{lemma}} mult\_transp\_ones\_sq\_mat:} \vspace*{0.1cm}\\  \texttt{\fontsize{17pt}{10pt}\selectfont
  \hspace*{.2cm}{\bf{assumes}} (A::real mat) $\in$ carrier\_mat z z}\vspace*{0.1cm} \\ \texttt{\fontsize{17pt}{10pt}\selectfont \hspace*{.2cm}{\bf{shows}} ($\mathbf{J}_{m}$ z)$^{T}$ * A = (0$_{m}$ z z)  $\longleftrightarrow$ A$^{T}$ * ($\mathbf{J}_{m}$ z) = (0$_{m}$ z z) \vspace{0.2cm}
  }} \\ 
\hline
\makecell{\hspace*{-.2cm} \fontsize{18pt}{10pt}\selectfont$
 $$ (L^{T}\mathbb{J}_{m})(i,j) = 
 (\mathcal{D}_{out} - \mathcal{D}_{in})(i,i)$$
$} &
  \makecell[l]{\\ \texttt{\fontsize{17pt}{10pt}\selectfont {\bf{lemma}} mult\_lap\_transp\_ones\_eq\_diff\_deg:} \vspace*{0.1cm}\\  \texttt{\fontsize{17pt}{10pt}\selectfont
  \hspace*{.01cm} 
   i\hspace*{-.1cm} $<$\hspace*{-.1cm} m $\Longrightarrow$ j\hspace*{-.1cm} $<$\hspace*{-.1cm} m $\Longrightarrow$} \texttt{\fontsize{17pt}{10pt}\selectfont \hspace*{.1cm}  
    ($\mathpzc{L}$$^{T}$ * $\mathbf{J}_{m}$\hspace*{-.1cm} m)\hspace*{-.1cm} $\$\$$\hspace*{-.1cm} (i,j)} \vspace{0.2cm} 
     \\ \texttt{\fontsize{17pt}{10pt}\selectfont \hspace*{.01cm}  
    = ($\mathpzc{D}$out)\hspace*{-.1cm} $\$\$$\hspace*{-.1cm} (i,i)-($\mathpzc{D}$in)}\hspace*{-.1cm} $\$\$$\hspace*{-.1cm} (i,i)}\vspace{0.2cm} 
   \\ 
\hline
\makecell{\hspace*{-.2cm} \fontsize{18pt}{10pt}\selectfont$
 $$ A = [a_{ij}] \in \mathbb{R}^{z\times z} \: \mathrm{is\: diagonal} $$ $\vspace*{.5cm} \\ \fontsize{18pt}{10pt}\selectfont$ $$\Longrightarrow (A\mathbb{J}_{m})(i,j) = A(i,i)$$
 $}
  &
  \makecell[l]{\\ \texttt{\fontsize{17pt}{10pt}\selectfont {\bf{lemma}} diag\_sq\_mult\_allone\_mat:} \vspace*{0.1cm}\\  \texttt{\fontsize{17pt}{10pt}\selectfont
  \hspace*{.2cm}{\bf{assumes}} diagonal\_mat (A::real mat) {\bf{and}} square\_mat A}\vspace*{0.1cm} \\ \texttt{\fontsize{17pt}{10pt}\selectfont \hspace*{.2cm}{\bf{shows}} i\hspace*{-.1cm} $<$\hspace*{-.1cm} dim\_row A $\Longrightarrow$ j\hspace*{-.1cm} $<$\hspace*{-.1cm} dim\_col A}  \vspace*{0.1cm} \\ \texttt{\fontsize{17pt}{10pt}\selectfont \hspace*{2.6cm} $\Longrightarrow$ (A * $\mathbf{J}_{m}$ (dim\_row A))\hspace*{-.1cm} $\$\$$\hspace*{-.1cm} (i,j) = A\hspace*{-.1cm} $\$\$$\hspace*{-.1cm} (i,i)\vspace{0.2cm}
 }} 
\\ \hline
\end{tabular}}
\end{center}
\caption{\centering \small Some Helper Lemmas for the Verification of Equation~(\ref{lemmacol})}
\label{tablemmacol}
\vspace*{-.6cm}
\end{table}
\\
\noindent Next, we verify the symmetry of the Laplacian matrix in the locale \texttt{sym\_wdg\_sys}. 
\vspace*{-.3cm}
\begin{flushleft}\label{symlap}
{\fontfamily{cmtt}\selectfont \small {\bf{lemma}} \ka{laplacian\_mat\_sym}:\hspace*{-.01cm} is\_symmetric $\mathpzc{L}$}
\end{flushleft}
\vspace*{-.1cm}
The above lemma indicates that the Laplacian matrix is symmetric if the edge weights are symmetric. Having formally defined and analyzed the Laplacian matrix derived from a weighted directed graph, we now formalize the abstract concept of the Laplacian, which applies to square matrices under certain conditions.
\begin{definition}
\label{DEF:isLAPMAT}Laplacian Matrix~\cite{bullo2018lectures} \\
\normalfont A real-valued square matrix $L$ is Laplacian if the following conditions are hold: 
\begin{itemize}
    \item[(1)] $L$ has non-positive non-diagonal entries.
    \item[(2)]  $L$ has non-negative diagonal entries.
    \item[(3)] The sum of each row of the matrix $L$ is zero.
\end{itemize}
\end{definition}

The formalization of the above definition is established as a predicate in Isabelle/HOL as follows:
\vspace*{-.1cm}
\begin{flushleft}
{\fontfamily{cmtt}\selectfont \small {\bf{definition}} 
\ka{is\_laplacian}\hspace*{-.1cm} ::\hspace*{-.1cm} real mat $\Rightarrow$\hspace*{-.1cm} bool
\\ \hspace*{.1cm}
{\bf{where}} \ka{is\_laplacian}\hspace*{-.1cm} L $\equiv$ square\_mat\hspace*{-.1cm} L \\ \hspace*{.2cm} $\land$  ($\forall$\hspace*{-.1cm} i\hspace*{-.1cm} $<$\hspace*{-.1cm} dim\_row\hspace*{-.1cm} L.\hspace*{-.1cm} $\forall$\hspace*{-.1cm} j\hspace*{-.1cm} $<$\hspace*{-.1cm} dim\_col L.\hspace*{-.1cm}
i\hspace*{-.1cm} $\neq$\hspace*{-.1cm} j\hspace*{-.1cm} $\longrightarrow$\hspace*{-.1cm} L\hspace*{-.1cm} $\$\$$\hspace*{-.1cm} (i,j)\hspace*{-.1cm} $\le$\hspace*{-.1cm} 0) \\ \hspace*{.2cm}
$\land$ ($\forall$\hspace*{-.1cm} i\hspace*{-.1cm} $<$\hspace*{-.1cm} dim\_row\hspace*{-.1cm} L.\hspace*{-.1cm} $\forall$\hspace*{-.1cm} j\hspace*{-.1cm} $<$\hspace*{-.1cm} dim\_col L.\hspace*{-.1cm}
i\hspace*{-.1cm} =\hspace*{-.1cm} j\hspace*{-.1cm} $\longrightarrow$\hspace*{-.1cm} L\hspace*{-.1cm} $\$\$$\hspace*{-.1cm} (i,j)\hspace*{-.1cm} $\ge$\hspace*{-.1cm} 0) \\ \hspace*{.2cm}
$\land$ ($\forall$\hspace*{-.1cm} i\hspace*{-.1cm} $<$\hspace*{-.1cm} dim\_row\hspace*{-.1cm} L.\hspace*{-.1cm} ($\sum$\hspace*{-.1cm} j\hspace*{-.1cm} $\in$\hspace*{-.1cm} $\{$0..$<$dim\_col L$\}$.\hspace*{-.1cm}
L\hspace*{-.1cm} $\$\$$\hspace*{-.1cm} (i,j)\hspace*{-.1cm} $=$\hspace*{-.1cm} 0)) 
}
\end{flushleft}
\vspace*{-.1cm}
We can then verify that the Laplacian matrix $\mathpzc{L}$ within the locale \texttt{wdg\_sys} satisfies the above conditions. 
\vspace*{-.1cm}
\begin{flushleft}\label{symmlap}
{\fontfamily{cmtt}\selectfont \small {\bf{lemma}} \ka{L\_laplacian}:\hspace*{-.01cm} is\_laplacian $\mathpzc{L}$}
\end{flushleft}
Here, it is trivial to verify the row-sum condition (condition 3 in Definition~\ref{DEF:isLAPMAT}) since we have already proven that $\mathpzc{L}$ is a square matrix whose row sums are zero. The other conditions are quickly verified with the properties, such as non-negative diagonal entries, non-positive off-diagonal entries, and the non-negativity of adjacency matrix entries. More details on the formalization and verification are available in~\cite{b25}. In the next section, we establish the relationships between the aforementioned network topology matrices and incidence matrices, which are also commonly used in network analysis.

\vspace*{-.2cm}
\section{Relationships between Network Topology Matrices
}
\label{sec6incirel}
\begin{figure}[htbp]
    \centering
    \includegraphics[width=1\linewidth]{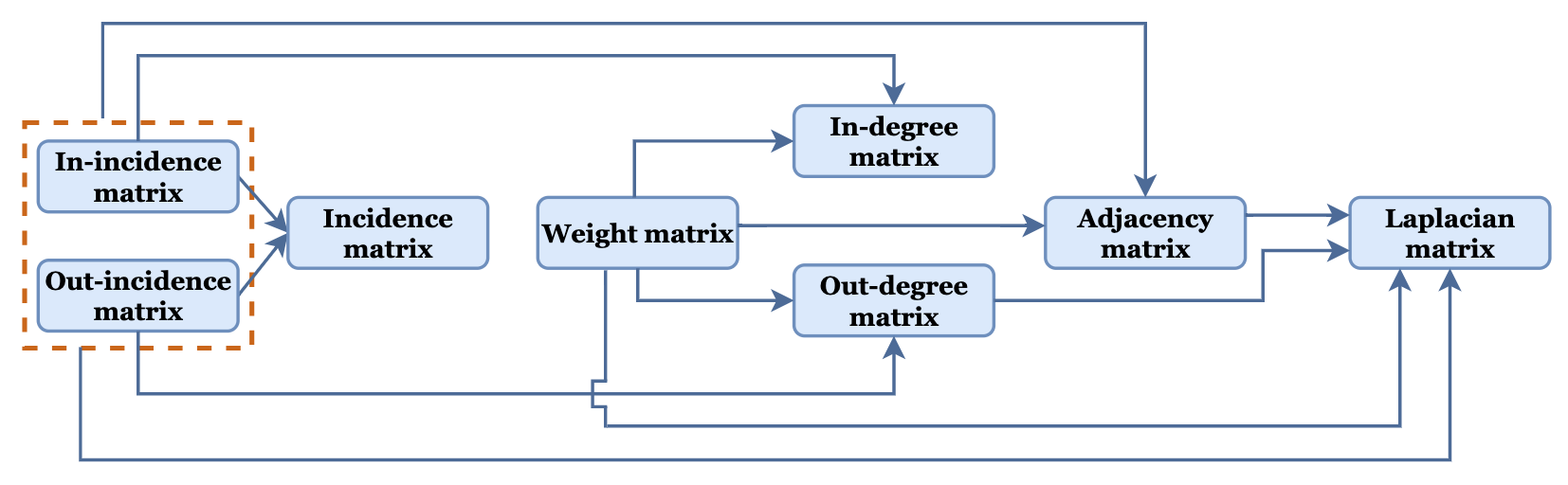}
    \caption{Relationships between Network Topology Matrices}
    \label{fig:relmatrices}
\vspace*{-.1cm}
\end{figure}

The Laplacian matrix is closely related to the adjacency and degree matrices, as discussed in Section~\ref{sec5adl}. Similarly, incidence matrices capture graph connectivity by defining relationships between nodes and edges. Establishing a connection between these representations offers a theoretically sound and comprehensive framework, which can be applicable to a broader range of systems. Figure~\ref{fig:relmatrices} illustrates the relationships between the network topology matrices of weighted directed graphs, with arrows indicating their interdependencies. For example, the Laplacian matrix is derived using the in-incidence and out-incidence matrices, along with the \textit{weight matrix}, which is a diagonal matrix representing edge weights~\cite{bullo2018lectures}. Similarly, the adjacency matrix can be obtained using the in-incidence, out-incidence and weight matrices.
 
 The incidence matrix and its basic properties were formalized in our prior work within the “\textit{Incidence\_Matrices}” theory~\cite{aksoy2025faecnttp}. However, to be able to derive the out-degree, in-degree, adjacency, and Laplacian matrices from the weight and incidence matrices, we have extended this framework by defining and formalizing the in-incidence and out-incidence matrices along with their associated properties. These additions are provided in the “\textit{Incidence\_Matrices\_Extras}” theory~\cite{b25}.

\vspace*{-.1cm}
\subsection{Incidence Matrices}

Incidence matrices represent network systems captured by directed graphs with no-self loops. These matrices are used to represent the relationships between nodes and edges without considering weights of the edges.  Consider a network with $m$ nodes, $n$ edges, and no self-loops, then we define the in-incidence and out-incidence matrices, which allow for a separate characterization of positive (outgoing) and negative (incoming) incidence relations. Their mathematical formulation is given in the following definition.
\vspace*{-.05cm}
\begin{definition}
\label{DEF:ininciout}
In-Incidence and Out-Incidence Matrices of a Directed Graph~\cite{bullo2018lectures} \\
 \normalfont An in-incidence matrix $\mathcal{I}_{in}$ and an out-incidence matrix $\mathcal{I}_{out}$ are defined respectively as $m \times n$ matrices: 
\begin{equation*}
 {[\mathcal{I}_{in}}]_{ij} = \left\{ \begin{array}{lll}
 1, & \textrm{if node $i$ equals to the tail of the edge $j$} \\ 
 0, & \textrm{otherwise} \\
\end{array} \right.   
\end{equation*}
\vspace*{-.2cm}
\begin{equation*}
 [{\mathcal{I}_{out}}]_{ij} = \left\{ \begin{array}{lll}
 1, & \textrm{if node $i$ equals to the head of the edge $j$} \\ 
 0, & \textrm{otherwise} \\
\end{array} \right.   
\end{equation*}
\noindent for $i \in \{1, 2, \dots , m\}$ and $j \in \{1,2, \dots , n\}$. 
\end{definition}

The Isabelle/HOL formalization of the out-incidence matrices is given as follows: 
\vspace*{-.1cm}
\begin{flushleft}
{\fontfamily{cmtt}\selectfont \small {\bf{definition}} \ka{out\_inc\_mat}\hspace*{-.1cm} ::\hspace*{-.1cm} $'$a\hspace*{-.1cm} nodes\hspace*{-.1cm} $\Rightarrow$\hspace*{-.1cm} $'$a\hspace*{-.1cm} edges\hspace*{-.1cm} $\Rightarrow$\hspace*{-.1cm} 
($'$b\hspace*{-.1cm} ::\hspace*{-.1cm} {ring\_1}\hspace*{-.1cm} mat) \\ \hspace*{.7cm}{\bf{where}}\hspace*{-.1cm} 
\ka{out\_inc\_mat}\hspace*{-.1cm} Ns\hspace*{-.1cm} Es $\equiv$ mat\hspace*{-.1cm} (length Ns)\hspace*{-.1cm} (length Es)\\ \hspace*{3.5cm}($\lambda$(i,j).\hspace*{-.1cm} if\hspace*{-.1cm} fst\hspace*{-.1cm} (Es!j)\hspace*{-.1cm} =\hspace*{-.1cm} (Ns!i)\hspace*{-.1cm} then\hspace*{-.1cm} 1\hspace*{-.1cm} else\hspace*{-.1cm} 0)
}
\end{flushleft}
\vspace*{-.01cm}

Here, we restrict the element type of the out-incidence matrix to a type $'$\texttt{b\hspace*{-.1cm} ::\hspace*{-.1cm} ring\_1}, which allows us to use the algebraic properties provided by the \texttt{ring\_1} type class when reasoning about its fundamental properties.
 
In a similar way, we also formalize in-incidence matrix in Isabelle/HOL, which is available at~\cite{b25}. We then formally verify this relationship within the locale \texttt{simple\_netw\_sys}, as follows:
\begin{flushleft}  
 {\fontfamily{cmtt}\selectfont \small {\bf{lemma}} \ka{incidence\_in\_out\_inc}:\hspace*{-.1cm} K\hspace*{-.1cm} =\hspace*{-.1cm} I\_out  -\hspace*{-.1cm} I\_in}
\end{flushleft}
Here, \texttt{I\_out} and \texttt{I\_in} denote \texttt{out\_inc\_mat\hspace*{-0.1cm} $\mathcal{N}$s\hspace*{-0.1cm} $\mathcal{E}$s} and \texttt{in\_inc\_mat\hspace*{-0.1cm} $\mathcal{N}$s\hspace*{-0.1cm} $\mathcal{E}$s} within the locale, respectively. Furthermore, we extend the theory of incidence matrix by verifying additional properties, such as uniqueness and the relationship between incidence matrices and vectors, which are available at~\cite{b25}.
\vspace*{-.1cm}
\subsection{Verification of the Matrices' Relationships}

We formalize the relationships between network topology matrices, as illustrated in Figure~\ref{fig:relmatrices}. We start by introducing a weight matrix, which is mathematically defined as follows:  
 \begin{definition}{\textit{Weight Matrix}}~\cite{bullo2018lectures} \\
\normalfont {The weight matrix $\mathcal{W} \in \mathbb{R}^{n\times n}$ of a weighted directed graph is a diagonal matrix defined as follows: 
\begin{equation*}
  \mathcal{W} = diag(\omega(e_{1}), \omega(e_{2}), \dots, \omega(e_{n}))   
\end{equation*} }
\end{definition}
\vspace*{-.2cm}
\noindent where $\{e_{1},e_{2},\dots, e_{n}\}$ is the set of edges in the graph. 

Using the weight matrix, the following fundamental relationships are established among the matrices. The out-degree and in-degree matrices are expressed in terms of their respective incidence matrices as: 
\begin{align} 
\mathcal{D}_{out} &=  \mathcal{I}_{out}\mathcal{W}\mathcal{I}_{out}^T \label{EQ: DOUTrel}
\end{align}
\vspace*{-1cm}
\begin{align}
 \mathcal{D}_{in} &=  \mathcal{I}_{in}\mathcal{W}\mathcal{I}_{in}^T\label{EQ: Dinrel}      
\end{align}

The adjacency matrix can be derived using the in-incidence, out-incidence and weight matrices, as expressed in the following equation:
\vspace*{-.01cm}
\begin{align}
\mathcal{A} &=  \mathcal{I}_{out}\mathcal{W}\mathcal{I}_{in}^T \label{EQ: adjrel}   
\end{align}
Finally, the relationship between Laplacian and incidence matrices is formulated as: 
\begin{align}
\mathcal{L} &=  \mathcal{I}_{out}\mathcal{W}(\mathcal{I}_{out} - \mathcal{I}_{in})^T = \mathcal{I}_{out}\mathcal{W}(\mathcal{I})^T \label{EQ:LAPrel}         
\end{align}

Notably, if a network system is modeled as an unweighted undirected graph, the weight matrix reduces to a square identity matrix. Now, to verify the generalized relationships between network topology matrices in Isabelle/HOL within the locale \texttt{simple\_wdg\_sys}, we first formalize the weight matrix as the following definition:
\vspace*{-0.01cm}
\begin{flushleft}
{\fontfamily{cmtt}\selectfont \small {\bf{definition}}
\ka{weight\_mat\_prop}\hspace*{-0.1cm} ::\hspace*{-0.1cm} real mat  $\Rightarrow$ bool {\bf{where}} \\ 
\hspace*{.1cm} \ka{weight\_mat\_prop} $\mathpzc{W}$ $\equiv$ diagonal\_mat $\mathpzc{W}$ $\land$ $\mathpzc{W}$\hspace*{-0.1cm} $\in$\hspace*{-0.1cm} carrier\_mat n n \\
\hspace*{1.2cm} $\land$ ($\forall$\hspace*{-0.1cm} i<dim\_row\hspace*{-0.1cm} $\mathpzc{W}$.\hspace*{-0.1cm} $\forall$\hspace*{-0.1cm} j<dim\_col\hspace*{-0.1cm} $\mathpzc{W}$.\hspace*{-0.1cm} i\hspace*{-0.1cm} =\hspace*{-0.1cm} j\hspace*{-0.1cm} $\longrightarrow$\hspace*{-0.1cm} $\mathpzc{W}$\hspace*{-0.1cm} $\$\$$\hspace*{-0.1cm} (i,j)\hspace*{-0.1cm} =\hspace*{-0.1cm} wei\hspace*{-0.1cm} ($\mathcal{E}$s!j)) }
\end{flushleft}
\vspace*{-0.01cm}
\noindent The above definition ensures that the weight matrix is a diagonal $n\times n$ matrix, where its diagonal elements correspond to the weights of the edges in order. We then verify Equation~(\ref{EQ: DOUTrel}) as the following theorem:
\vspace*{-0.01cm}
\begin{flushleft}
{\fontfamily{cmtt}\selectfont \small {\bf{lemma}}
\ka{rel\_D\_out\_I\_out}:\\
\hspace*{.2cm} {\bf{assumes}} weight\_mat\_prop\hspace*{-0.1cm} $\mathpzc{W}$ {\bf{shows}} D\_out\hspace*{-0.1cm} =\hspace*{-0.1cm} I\_out\hspace*{-0.1cm} *\hspace*{-0.1cm} $\mathpzc{W}$\hspace*{-0.1cm} *\hspace*{-0.1cm} (I\_out)$^{T}$}
\end{flushleft}

\noindent The theorem is verified using \texttt{eq\_matI}, where dimensional compatibility is discharged automatically. The proof proceeds by establishing element-wise equality, distinguishing between diagonal and off-diagonal entries. The off-diagonal case follows from structural properties of the network, while the diagonal case is obtained through algebraic manipulation to match the definition of the out-degree matrix. Next, we verify Equation~(\ref{EQ: Dinrel}) in Isabelle/HOL as follows: 

\begin{flushleft}
{\fontfamily{cmtt}\selectfont \small {\bf{lemma}}
\ka{rel\_D\_in\_I\_in}:\\
 \hspace*{.2cm} {\bf{assumes}} weight\_mat\_prop\hspace*{-0.1cm} $\mathpzc{W}$ {\bf{shows}} D\_in\hspace*{-0.1cm} =\hspace*{-0.1cm} I\_in\hspace*{-0.1cm} *\hspace*{-0.1cm} $\mathpzc{W}$\hspace*{-0.1cm} *\hspace*{-0.1cm} (I\_in)$^{T}$}
\end{flushleft}

\noindent The verification of the above lemma is very similar to that of lemma \texttt{rel\_D\_out\\\_I\_out}. More details
about its verification can be found in~\cite{b25}. Next, we verify the relationship between the adjacency matrix, and the in-incidence and out-incidence matrices (Equation~(\ref{EQ: adjrel})) as follows:

\begin{flushleft}
{\fontfamily{cmtt}\selectfont \small {\bf{lemma}}
\ka{rel\_adj\_Iout\_Iin}:\\
\hspace*{.1cm} {\bf{assumes}} weight\_mat\_prop\hspace*{-0.1cm} $\mathpzc{W}$ {\bf{shows}}\hspace*{-0.1cm} $\mathpzc{A}$\hspace*{-0.1cm} =\hspace*{-0.1cm} I\_out\hspace*{-0.1cm} *\hspace*{-0.1cm} $\mathpzc{W}$\hspace*{-0.1cm} *\hspace*{-0.1cm} (I\_in)$^{T}$}
\end{flushleft}

The verification of the  theorem follows a nested case analysis strategy similar to earlier proofs. The main challenge arises in the off-diagonal case, which depends on whether an edge exists between two nodes. When an edge exists, we identify its unique index and prove that the matrix product isolates the corresponding term, while all other terms vanish using the indexing properties of the incidence matrices and basic summation reasoning. When no edge exists, both the adjacency entry and the matrix corresponding product becomes zero. This proof requires explicitly tracking edges and their positive and negative incident nodes. Our list-based representation simplifies this process, as it allows us to work directly with indices without maintaining separate mappings between nodes/edges and their positions.

Finally, we verify Equation~(\ref{EQ:LAPrel}) in Isabelle/HOL as follows: 
\begin{flushleft}
{\fontfamily{cmtt}\selectfont \small {\bf{lemma}}
\ka{rel\_lap\_inc}:\\
\hspace*{.1cm} {\bf{assumes}} weight\_mat\_prop\hspace*{-0.1cm} $\mathpzc{W}$ \\
\hspace*{.1cm} {\bf{assumes}} $\bigwedge$i.\hspace*{-0.1cm} i<\hspace*{-0.1cm} m\hspace*{-0.1cm} $\Longrightarrow$\hspace*{-0.1cm} wei\_outdeg\hspace*{-0.1cm} ($\mathcal{N}$s!i)\hspace*{-0.1cm} =\hspace*{-0.1cm} wei\_outdegree\hspace*{-0.1cm} ($\mathcal{N}$s!i) \\
\hspace*{.1cm} {\bf{shows}} $\mathpzc{L}$\hspace*{-0.1cm} = I\_out\hspace*{-0.1cm} *\hspace*{-0.1cm} $\mathpzc{W}$\hspace*{-0.1cm} *\hspace*{-0.1cm} (K)$^{T}$}
\end{flushleft}

The second assumption ensures the equivalence between the implicit and explicit formulations of the weighted out-degree of a node. The verification of the above theorem relies on previously verified theorems \texttt{rel\_adj\_Iout\_Iin} and \texttt{rel\_D\_out\_I\_out}, along with reasoning over matrix operations. 

This section establishes the formal relationships among the topological matrices, ensuring their consistency within the framework. In particular, the Laplacian matrix emerges as a unifying representation, motivating its usage in the applications, which we further explore in the next section.

 \section{Applications} 
\label{sec7app}
\vspace*{-.1cm}
In this section, we present two case studies that illustrate the applicability of our formalization. First, we apply our framework on Kron reduction, which is an algebraic method frequently used in network analysis within the domains of electrical and power engineering. Second, we formally analyze an electrical resistive network in terms of the power dissipation across its resistors. 
\vspace*{-.1cm}
\subsection{Kron Reduction of the Laplacian Matrix}

Kron reduction~\cite{kron1939tensor} is a widely used algebraic technique for analyzing complex and large-scale networks. It simplifies a network model by eliminating a subset of interior nodes from its corresponding graph, while preserving the essential topological characteristics of the original network. In the literature, this method has been applied to various networks such as electrical networks, smart grids and chemical reaction networks. For example, in~\cite{dorfler2012kron}, Kron reduction is applied to electrical networks captured by undirected graphs and represented algebraically by the Laplacian matrices. The generalized Kron reduction method for directed graphs is analyzed in~\cite{sugiyama2023kron}. Moreover, the Kron reduction method plays a crucial role in several theoretical problems in stochastic processes, spectral graph theory and knot theory~\cite{meyer1989stochastic,perk2006yang}.

\begin{figure}[htbp]
  \subcaptionbox*{\hspace*{-.6cm}(a) Wiring Diagram of Area 4}[.37\linewidth]{%
   \hspace*{-.31cm} \includegraphics[width=.91\linewidth]{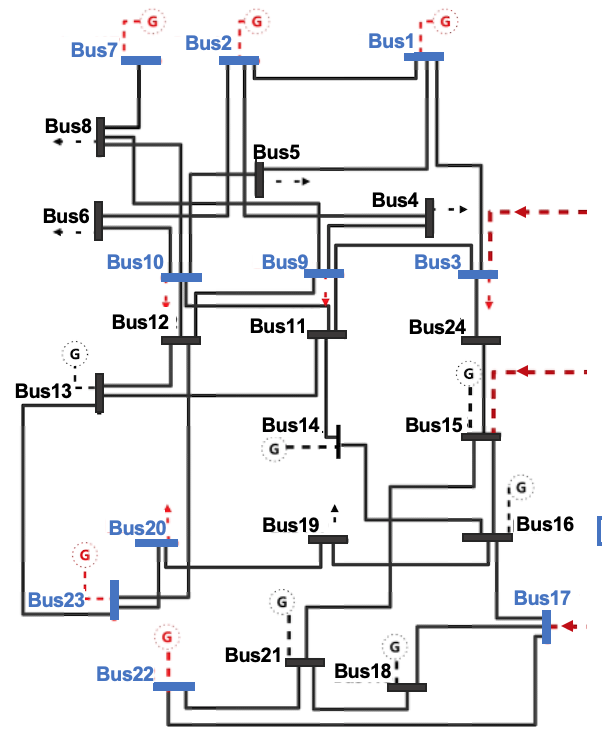}%
     \label{Fig: rts1}
  }%
  \hfill
  \subcaptionbox*{\hspace*{-.7cm}(b) Graph Representation}[.3\linewidth]{%
    \hspace*{-.31cm}\includegraphics[width=.91\linewidth]{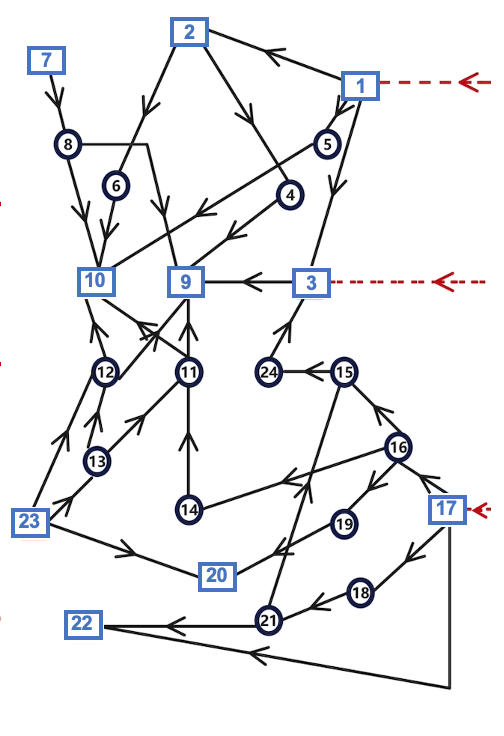}%
     \label{Fig: rtsg}
     }%
  \hfill
  \subcaptionbox*{\hspace*{-.6cm}(c) Kron Reduction of (b)}[.32\linewidth]{%
    \hspace*{-.31cm}\includegraphics[width=1\linewidth]{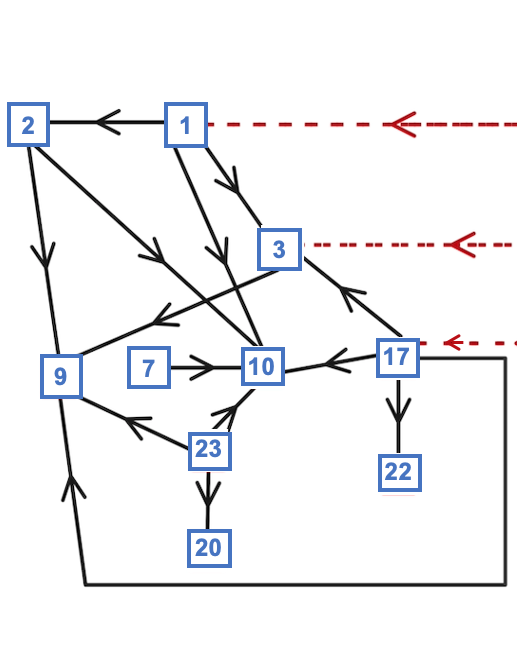}%
     \label{Fig: kr-rts}
  }
  {\caption{Area 4 of the Modified IEEE RTS-96 Test System and Kron reduction of its Graph Representation~\cite{wang2023modelling}}
  \label{Fig: kr-ieee}}
\end{figure}

Figure~\ref{Fig: kr-ieee} illustrates an area of the modified IEEE Reliability Test System (IEEE RTS-96)~\cite{tosatto2019modified}, which is widely used for evaluating power system reliability. It is a multi-area system consisting of 4 areas, each representing a 24-bus system with 24 nodes (buses and interior nodes) and 33 weighted edges. Here, to simplify the illustration, the edge weights are omitted in Figure~\ref{Fig: kr-ieee}. Applying the Kron reduction to the Laplacian matrix corresponding to the graph in Figure~\ref{Fig: kr-ieee}{\ka{(b)}}, yields a Kron-reduced Laplacian matrix that retains the essential topological characteristics of the original network.
\vspace*{-.4cm}
\vspace*{-.5cm}
\par We now mathematically define the generalized Kron reduction as follows:
\begin{definition}
\label{DEF:KR}
Kron Reduction of Weighted Directed Graph~\cite{sugiyama2023kron} \\
\normalfont Consider a network captured by a weighted directed graph with $m$ nodes and self-loops. Let $\alpha$ be a proper subset of nodes such that $|\alpha| \ge 2$. The corresponding Laplacian matrix $\mathcal{L}$ can be expressed as a block matrix of the form:
\begin{equation}    
\label{eq:lblok}
  \mathcal{L}_{m\times m} = \left[\begin{array}{ c  c }
    \mathcal{L}_{|\alpha|\times |\alpha|} & \mathcal{L}_{|\alpha|\times |\beta|} \\    
    \mathcal{L}_{|\beta|\times |\alpha|} & \mathcal{L}_{|\beta| \times |\beta|}
  \end{array}\right]
\end{equation}
where $\alpha \cup \beta = \{1,\dots,m\}$ and $\beta = \alpha^{c}$. 

\par A Kron-reduced Laplacian denoted by $\mathcal{L}_{red}$, is defined as: 
\begin{equation}
\label{eq:lred}
 \mathcal{L}_{red} := \mathcal{L}\:[\alpha, \alpha] -  \mathcal{L}\:[\alpha, \beta]  \mathcal{L}\:[\beta, \beta]^{-1}  \mathcal{L}\:[\beta, \alpha] 
\end{equation}
\end{definition}

\noindent Here, $\beta$ is the complement of the boundary node set ($\alpha$), which corresponds to the interior (eliminated) nodes. In addition, $\mathcal{L}\:[\beta, \beta]$ is an invertible and non-negative submatrix of $\mathcal{L}$, which is obtained by the rows and columns indexed using $\beta$. We now formalize the Kron reduction in Isabelle/HOL as follows:

\begin{flushleft}
{\fontfamily{cmtt}\selectfont \small {\bf{definition}} 
\ka{kron\_red}\hspace*{-.1cm} ::\hspace*{-.1cm} $'$b\hspace*{-.1cm} ::\hspace*{-.1cm} field mat $\Rightarrow$\hspace*{-.1cm} $'$a list $\Rightarrow$\hspace*{-.1cm} $'$a list $\Rightarrow$\hspace*{-.1cm} $'$b mat
\\ \hspace*{.1cm}
{\bf{where}} \ka{kron\_red}\hspace*{-.1cm} M\hspace*{-.1cm} N1\hspace*{-.1cm} N2\hspace*{-.1cm} $\equiv$\hspace*{-.1cm} (\ka{let}\hspace*{-.1cm} a\hspace*{-.1cm} =\hspace*{-.1cm} length\hspace*{-.1cm} N1;\hspace*{-.1cm} b\hspace*{-.1cm} =\hspace*{-.1cm} length N2; \\ \hspace*{.4cm}
\hspace*{-.1cm} split\_block\hspace*{-.1cm} M\hspace*{-.1cm} a\hspace*{-.1cm} b\hspace*{-.1cm} =\hspace*{-.1cm} (A,B,C,D)\hspace*{-.1cm} \ka{in}\hspace*{-.1cm} (A\hspace*{-.1cm} -\hspace*{-.1cm} B\hspace*{-.1cm} *\hspace*{-.1cm} (the\hspace*{-.1cm} (mat\_inverse\hspace*{-.1cm} D))\hspace*{-.1cm} *\hspace*{-.1cm} C)) 
}
\end{flushleft}
\noindent where \texttt{split\_block} is a function that decomposes a matrix \texttt{M} into four submatrices based on the specified row and column indices, \texttt{a} and \texttt{b}, respectively. Several useful results are obtained from this definition, which are further used to verify the properties of Kron-reduced Laplacian matrix. Accordingly, we verify the following lemma within the locale \texttt{partitioned\_wdg\_sys}  to simplify the Kron reduction definition:

\begin{flushleft}
{\fontfamily{cmtt}\selectfont \small {\bf{lemma}} \ka{kr\_lap\_pred\_exp}: \\
\hspace*{.2cm} {\bf{assumes}} kr\_lap\_pred A B C D D$'$ \\
\hspace*{.56cm} {\bf{shows}} $\mathpzc{L}$ = four\_block\_mat A B C D \\
\hspace*{.9cm} {\bf{and}} (kron\_red $\mathpzc{L}$ N1 N2)\hspace*{-.1cm} $\in$\hspace*{-.1cm} carrier\_mat\hspace*{-.1cm} (length N1)\hspace*{-.1cm} (length N1)\\
\hspace*{.9cm} {\bf{and}} kron\_red $\mathpzc{L}$ N1 N2 = A - B * D$'$ * C \\
\hspace*{.9cm} {\bf{and}} A\hspace*{-.1cm} $\in$\hspace*{-.1cm} carrier\_mat\hspace*{-.1cm} (length N1)\hspace*{-.1cm} (length N1) \\
\hspace*{.9cm} {\bf{and}} B\hspace*{-.1cm} $\in$\hspace*{-.1cm} carrier\_mat\hspace*{-.1cm} (length N1)\hspace*{-.1cm} (length N2) \\
\hspace*{.9cm} {\bf{and}} C\hspace*{-.1cm} $\in$\hspace*{-.1cm} carrier\_mat\hspace*{-.1cm} (length N2)\hspace*{-.1cm} (length N1) \\
\hspace*{.9cm} {\bf{and}} D\hspace*{-.1cm} $\in$\hspace*{-.1cm} carrier\_mat\hspace*{-.1cm} (length N2)\hspace*{-.1cm} (length N2) 
}
\end{flushleft}

\noindent Here, the locale characterizes that the node list ($\mathcal{N}$s) is partitioned into two proper sublists (\texttt{N1} and \texttt{N2}), which are both nonempty and distinct. Moreover, the assumption ensures that: (1) the matrix \texttt{D} has an inverse \texttt{D}$'$ of size (\texttt{length N2}) $\times$ (\texttt{length N2}), and (2) the Laplacian matrix $\mathpzc{L}$ can be partitioned into submatrices \texttt{A,B,C,D} at specified indices for rows (\texttt{length N1}) columns (\texttt{length N2}), corresponding to the Equation~(\ref{eq:lblok}). 

The Kron reduction satisfies the closure property, meaning that the Kron-reduced Laplacian matrix is itself a Laplacian matrix. This follows from the fact that the Kron reduction preserves the topological structure of the network, its algebraic manipulation does not alter the fundamental connectivity of the system. We verify this property in the following Isabelle/HOL lemma:

\begin{flushleft}
{\fontfamily{cmtt}\selectfont \small {\bf{lemma}} \ka{Lred\_is\_Laplacian}: \\
\hspace*{.2cm} {\bf{assumes}} kr\_lap\_pred A B C D D$'$ \\
\hspace*{.2cm} {\bf{and}} $\bigwedge$i\hspace*{-0.1cm} j.\hspace*{-0.1cm} i$<$\hspace*{-0.1cm} length N2\hspace*{-0.1cm} $\Longrightarrow$\hspace*{-0.1cm} j\hspace*{-0.1cm} $<$\hspace*{-0.1cm} length\hspace*{-0.1cm} N2\hspace*{-0.1cm} $\Longrightarrow$\hspace*{-0.1cm} D$'$\hspace*{-0.1cm} $\$\$$\hspace*{-0.1cm} (i,j)\hspace*{-0.1cm} $\ge$\hspace*{-0.1cm} 0 \\
\hspace*{.2cm} {\bf{show}} is\_laplacian (kron\_red $\mathpzc{L}$ N1 N2) 
}
\end{flushleft}
The proof involves verifying the three conditions specified in Definition~\ref{DEF:isLAPMAT}, using Laplacian properties such as matrix dimensionality, zero row-sum, and element indexing. Additionally, we establish new identities for block matrices, analyze the sign of each block’s entries, and prove the homogeneous block matrix–vector equation. Through this verification, we formally confirm the closure property, demonstrating that a reduced network system can be reliably replaced by its Kron-reduced equivalent, thereby supporting trustworthy model reduction. To the best of our knowledge, there exists no formalization of Kron reduction for graphs in any interactive theorem provers.

\subsection{Power Dissipation in Resistive Circuits}

\begin{figure}[htbp]
    \centering
    \includegraphics[width=1\linewidth]{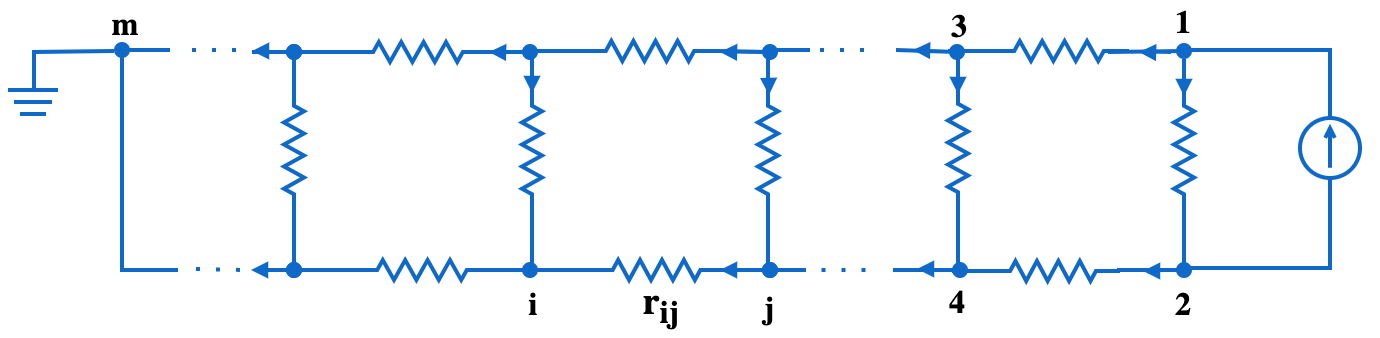}
    \caption{Resistive Electrical Network}
    \label{fig:rescirc}
\end{figure}

The power dissipation in an electrical network is a fundamental concept that determines the system’s overall efficiency and reliability. In this subsection, we verify the power dissipation for a purely resistive electrical network using the Laplacian matrix. A resistive electrical network is composed of resistors and voltage or current sources. As a circuit element, a resistor plays a vital role in limiting the current flow and controlling the system's functionality by maintaining an appropriate amount of current. Consider a resistive electrical network as depicted in Figure~\ref{fig:rescirc} with $m$ nodes. This network can be abstractly represented by a weighted directed graph, where each edge represents a resistor. For instance, an edge $r_{ij}$ describes a resistor, and the current flow in the resistor is directed from node $j$ to node $i$. 
\vspace*{-.3cm}
\vspace*{-.4cm}
\par For an arbitrary edge, represented by a pair of nodes, Ohm's law~\cite{monier2001electric} through the edge is mathematically expressed as follows:

\begin{equation}
\label{cij}
    c_{ij} = \frac{(v_{i} - v_{j})}{r_{ij}} = a_{ij}(v_{i} - v_{j})
\end{equation}

\noindent where $c_{ij}$ represents the current flowing from node $i$ to node $j$, and $v_{i}$ and $v_{j}$ correspond to the voltages at the nodes $i$ and node $j$, respectively. Furthermore, $a_{ij}$ represents the conductance of the circuit, which is defined as the inverse of the resistance, i.e., $a_{ij} = \frac{1}{r_{ij}}$. If there is no edge from node $i$ to node $j$, then $a_{ij}$ is set as 0. Moreover, we assume that $r_{ij}>0$. The injected current at node $i$ is modeled via KCL and Ohm's law as follows:

\begin{equation}
\label{cinj}
    c_{\textit{inj}}  = \sum_{j =1, j \neq i}^{m} (c_{ij}) = 
\sum_{j =1, j \neq i}^{m} a_{ij}(v_{i}-v_{j})
\end{equation}

The total power dissipation in the electrical network is given, as follows: 

\begin{equation}
\label{power}
    P_{dissipated} = \mathbf{v}^T\mathcal{L}\:\mathbf{v}
\end{equation}

\noindent where {$\mathbf{v} = [v_{1}, v_{2},\dots,v_{m}]^{T}$} represents the vector of voltages at nodes, and $\mathcal{L}$ refers to the Laplacian matrix of given network. We now formalize the current flowing between node $i$ and node $j$, which describes Ohm's law along the corresponding edge (Equation~(\ref{cij})), as follows:

\begin{flushleft}
{\fontfamily{cmtt}\selectfont \small {\bf{definition}}
\ka{cur\_flow\_btw\_nodes} {\bf{where}} \\ \hspace*{1cm} \ka{cur\_flow\_btw\_nodes}\hspace*{-0.1cm} v\hspace*{-0.1cm} i\hspace*{-0.1cm} j\hspace*{-0.1cm} $\equiv$\hspace*{-0.1cm} $\mathpzc{A}$\hspace*{-0.1cm} $\$\$$\hspace*{-0.1cm} (i,j)\hspace*{-0.1cm} *\hspace*{-0.1cm} (v$\$$i\hspace*{-0.1cm} -\hspace*{-0.1cm} v$\$$j)} 
\end{flushleft}

\noindent The function \texttt{cur\_flow\_btw\_nodes} accepts the vector of voltages \texttt{v} and the indices $i$ and $j$ as inputs, and returns the result of the multiplication of the voltage difference and conductance. Here, the conductance is modeled as an entry of the adjacency matrix. Next, we formalize the sum of current flows (using KCL) for each node $i$, which gives the injected current at node $i$ (Equation~(\ref{cinj})) as:

\begin{flushleft}
{\fontfamily{cmtt}\selectfont \small 
{\bf{definition}}
\ka{KCL\_cinj}\hspace*{-0.1cm} ::\hspace*{-0.1cm} real vec $\Rightarrow$ nat $\Rightarrow$ real {\bf{where}} \\ \hspace*{1cm} \ka{KCL\_cinj}\hspace*{-0.1cm} v\hspace*{-0.1cm} i\hspace*{-0.1cm} $\equiv$\hspace*{-0.1cm} ($\sum$\hspace*{-0.1cm} j\hspace*{-0.1cm} $\in$\hspace*{-0.1cm} ($\{$0..$<$\hspace*{-0.1cm} m$\}$\hspace*{-0.1cm} -\hspace*{-0.1cm} $\{$i$\}$).\hspace*{-0.1cm} cur\_flow\_btw\_nodes\hspace*{-0.1cm} v\hspace*{-0.1cm} i\hspace*{-0.1cm} j)}
\end{flushleft}

\noindent Here, we exclude the case ``\texttt{i=j}'' since it does not contribute to the summation. In the next step, we verify that the vector of injected currents and the voltages at the nodes $v_{i}$ satisfy the following equality, which is formalized as:

\begin{flushleft}
{\fontfamily{cmtt}\selectfont \small {\bf{lemma}}
\ka{lap\_c\_inj}:\\
\hspace*{.1cm} {\bf{assumes}} v\hspace*{-0.1cm} $\in$\hspace*{-0.1cm} carier\_vec\hspace*{-0.1cm} m\hspace*{-0.1cm} {\bf{and}}\hspace*{-0.1cm} i\hspace*{-0.1cm} $<$\hspace*{-0.1cm} dim\_vec\hspace*{-0.1cm} v \\
\hspace*{.1cm} {\bf{shows}} KCL\_cinj\hspace*{-0.1cm} v\hspace*{-0.1cm} i\hspace*{-0.1cm} =\hspace*{-0.1cm} ($\mathpzc{L}$ *$_{v}$ v)\hspace*{-0.1cm} $\$$\hspace*{-0.1cm} i}
\end{flushleft}

\noindent The above lemma provides an alternative formulation of Equation~(\ref{cinj}) using the Laplacian matrix. Its verification primarily relies on the following equality: 

\[(\mathcal{L}\mathbf{v})_{i} = \sum_{j=1,j\neq i}^{m} a_{ij}(v_{i} - v_{j}) \quad \forall\:\mathbf{v} \in \mathbb{R}^{m}\]

 We verify the above equality in Isabelle/HOL as follows:

\begin{flushleft}
{\fontfamily{cmtt}\selectfont \small {\bf{lemma}} \ka{useful\_eq1}:\\ 
\hspace*{.1cm} {\bf{assumes}} v $\in$ carrier\_vec m \\ 
\hspace*{.1cm} 
 {\bf{shows}} i\hspace*{-.1cm} $<$\hspace*{-.1cm} dim\_vec\hspace*{-.1cm} v\hspace*{-.1cm} $\Longrightarrow$ \\ \hspace*{2.2cm} ($\mathpzc{L}$ \hspace*{-.1cm} *$_{v}$\hspace*{-.1cm} v)\hspace*{-.1cm} $\$$\hspace*{-.1cm} i\hspace*{-.1cm} = \hspace*{-.1cm}
 ($\sum$\hspace*{-.1cm} j\hspace*{-.1cm} $\in$\hspace*{-.1cm} $\{$j.\hspace*{-.1cm} j\hspace*{-.1cm} $<$\hspace*{-.1cm} m\hspace*{-.1cm} $\land$\hspace*{-.1cm} j\hspace*{-.1cm} $\neq$\hspace*{-.1cm} i$\}$.\hspace*{-.1cm} $\mathpzc{A}$\hspace*{-.1cm} $\$\$$\hspace*{-.1cm} (i,j)\hspace*{-.1cm} *\hspace*{-.1cm} (v$\$$i\hspace*{-.1cm} -\hspace*{-.1cm} v$\$$j))} 
 
\end{flushleft}

\noindent The above lemma is verified by leveraging the indexing and dimension properties of the Laplacian matrix, along with some reasoning on sets and summations. Subsequently, the power dissipated on an edge (i.e., a resistor) is formalized as follows: 

\begin{flushleft}
{\fontfamily{cmtt}\selectfont \small 
{\bf{definition}}
\ka{power\_dissipated} {\bf{where}} \\ \hspace*{.1cm} \ka{power\_dissipated}\hspace*{-0.1cm} v\hspace*{-0.1cm} i\hspace*{-0.1cm} j $\equiv$\hspace*{-0.1cm} 
(cur\_flow\_btw\_nodes\hspace*{-0.1cm} v\hspace*{-0.1cm} i\hspace*{-0.1cm} j)$^{2}$ * (1/$\mathpzc{A}$}\hspace*{-.1cm} $\$\$$\hspace*{-.1cm} (i,j))
\end{flushleft}

The above definition mimics the power computation based on Ohm's law, i.e., $P = I^{2}R$, where $I$ refers to the current and $R$ represents the resistance of a network. The total power dissipated across the resistors is independent of the direction of the current flow. Hence, we verify the total power dissipation of the resistive electrical network within the locale \texttt{sym\_wdg\_sys}, as follows: 

\begin{flushleft}
{\fontfamily{cmtt}\selectfont \small {\bf{lemma}}
\ka{total\_pow\_dissp}:\\
\hspace*{.1cm} {\bf{assumes}} v\hspace*{-0.1cm} $\in$\hspace*{-0.1cm} carier\_vec\hspace*{-0.1cm} m\hspace*{-0.1cm}  \\
\hspace*{.1cm} {\bf{shows}}\hspace*{-0.1cm} 1/2\hspace*{-0.1cm} *\hspace*{-0.1cm} ($\sum$\hspace*{-0.1cm} i\hspace*{-0.1cm} $\in$\hspace*{-0.1cm} $\{$0..$<$m$\}$.\hspace*{-0.1cm} $\sum$\hspace*{-0.1cm} j\hspace*{-0.1cm} $\in$\hspace*{-0.1cm} $\{$0..$<$m$\}$.\hspace*{-0.1cm} power\_dissipated\hspace*{-0.1cm} v\hspace*{-0.1cm} i\hspace*{-0.1cm} j)\hspace*{-0.1cm} \\
\hspace*{8cm} =\hspace*{-0.1cm} inner\_prod\hspace*{-0.1cm} v\hspace*{-0.1cm} ($\mathpzc{L}$} *$_{v}$\hspace*{-0.1cm} v)
\end{flushleft}

\noindent where the function \texttt{inner\_prod} takes two vectors as input and returns the product of the transpose of the first vector with the second vector. The verification of the above lemma is based on the definition \texttt{power\_dissipated} and the following equality: 

\[\mathcal{L} = \mathcal{L}^T \Longrightarrow \mathbf{v}^{T}\mathcal{L}\:\mathbf{v} = \frac{1}{2}\sum_{i,j=1}^{m} a_{ij}(v_{i} - v_{j})^{2}\]

\noindent  We utilize some results on inner product in~\cite{liu2019formal} to establish this equation, and it is formally verified as the following lemma:

\begin{flushleft}
{\fontfamily{cmtt}\selectfont \small {\bf{lemma}} \ka{useful\_eq2}:\\ 
\hspace*{.1cm} {\bf{assumes}} v $\in$ carrier\_vec m \\ 
\hspace*{.1cm} 
 {\bf{shows}} inner\_prod\hspace*{-.1cm} v\hspace*{-.1cm} ($\mathpzc{L}$\hspace*{-.1cm} *$_{v}$\hspace*{-.1cm} v)\hspace*{-.1cm} \\ 
 \hspace*{2cm}= 1/2\hspace*{-0.1cm} *\hspace*{-0.1cm} ($\sum$\hspace*{-0.1cm} i\hspace*{-0.1cm} $\in$\hspace*{-0.1cm} $\{$0..$<$m$\}$.\hspace*{-0.1cm} $\sum$\hspace*{-0.1cm} j\hspace*{-0.1cm} $\in$\hspace*{-0.1cm} $\{$0..$<$m$\}$.\hspace*{-0.1cm} $\mathpzc{A}$\hspace*{-.1cm} $\$\$$\hspace*{-.1cm} (i,j)\hspace*{-.1cm} *\hspace*{-.1cm} (v\hspace*{-0.1cm} $\$$\hspace*{-.1cm} i\hspace*{-0.1cm} -\hspace*{-.1cm} v$\$$\hspace*{-.1cm} j)$^{2}$)}
 
\end{flushleft}
\vspace*{-.1cm}
\noindent The above lemma is proven based on the generalization of lemma \texttt{useful\_eq1}, i.e., summation is taken over $m$, and  \texttt{laplacian\_mat\_sym} (presented in Section~\ref{sec5lap}), along with some set, summation and arithmetic reasoning. This concludes the formal verification of the total power dissipation in resistive circuits, using the Laplacian matrix, demonstrating the applicability of the proposed theorem proving-based approach in reasoning about network topology matrices. More details about the above formalization and proof processes can be found in~\cite{b25}. 
\vspace*{-.4cm}
\section{Discussion}\label{SEC:Dissc}
\vspace*{-.1cm}
We have successfully formalized the network topology matrices, including the adjacency, degree, Laplacian and incidence matrices of weighted directed graphs and verified their properties in Isabelle/HOL. To the best of our knowledge, these formalizations are not available in other theorem provers. Our development ensures the correctness of the underlying structure by rigorously verifying the relationships between these matrices. In particular, we have established the inherent connection between graphs and matrices, where matrices are derived from network system parameters (e.g., nodes $\mathcal{N}$s and edges $\mathcal{E}$s). \\
\indent The formalization is generic, making the theorems and lemmas applicable to any finite number of nodes and edges (e.g., $i=1,2,\dots,m$ and $j=1,2, \dots,n$ denote indices for nodes and edges, respectively). Such generality is typically not achievable through simulation-based analysis of network topology matrices. One of the main challenges in this formalization was the informal and intuitive nature of some proofs in the literature. For instance, while the proof of the relationship between the adjacency and incidence matrices is typically condensed to five lines in textbooks (e.g.,~\cite{bullo2018lectures}), our Isabelle/HOL proofs (see the lemma \texttt{rel\_adj\_Iout\_\\Iin} in Section~\ref{sec6incirel}) required approximately 73 lines of code. Although this approach demands more effort, it ensures rigorous reasoning, capturing every intermediate step, thereby achieving a level of precision that is often overlooked in traditional mathematical treatments. \\
\indent Furthermore, in this paper, we established the simpler equivalence of the matrices and verified classical and structural properties within an appropriate locale. For some fundamental lemmas involving indexing and dimensional properties, we leveraged the powerful Sledgehammer automation tool~\cite{sled} in Isabelle/HOL, which allows the derivation of concise one-line proofs. Theorems such as \texttt{rel\_adj\_Iout\_Iin} and \texttt{rel\_lap\_inc} are proven under locale assumptions encapsulated within \texttt{simple\_wdg\_sys}, eliminating the need for repetitive assumptions in theorem statements. This modeling strategy not only reduces redundancy but also enhances modularity and extensibility, allowing for seamless integration with previous or future formalizations. Another example is that we established new locales, such as \texttt{wdg\_sys} and \texttt{sym\_netw\_sys}, by utilizing the locale inheritance from our prior network system formalization~\cite{aksoy2025faecnttp}. \\
\indent The formalizations developed in this work were designed to be as generic and reusable as possible, given the broad applicability of network topology matrices (as discussed in Section~\ref{Sec:Intro}). To this end, we first constructed the matrices outside the locales and then addressed their simpler instances within the locales. Although this approach may initially seem restrictive, integrating the matrices within network system locales is a necessary trade-off to achieve generality. For example, complex-valued network topology matrices can be readily defined by instantiating the matrix type as \texttt{complex} within the locale, thereby enhancing reusability and eliminating redundancy outside the locale. \\
\indent We adopted this design choice (similiar to~\cite{b21}) primarily to address challenges in defining the Laplacian matrix over a field within the locale \texttt{wdg\_sys}. Specifically, it is not feasible to instantiate both real-valued and complex-valued Laplacian matrices simultaneously, as complex numbers cannot be ordered in a way consistent with their algebraic properties. Instead of providing multiple definitions of the Laplacian matrix, one may establish a dedicated locale for complex network systems and prove the equivalence of the definitions within that locale. Accordingly, network systems formalized through a polymorphic type $'$\texttt{a}, supporting extensions, such as subgraphs, in which node and index types (e.g., \texttt{nat}) may differ. While this abstraction may increase proof complexity, Isabelle/HOL’s interpretation mechanism would be an effective way to instantiate locale parameters and manage this complexity. Moreover, adopting a generic node type facilitates the future integration of this formalization with other Isabelle/HOL libraries, many of which follow a similar parametric design.\\
 \indent We also presented practical applications to illustrate the usability of our formalization. For instance, we formalized the Kron reduction algebraic method for Laplacian matrices. Since the matrix formalizations in this paper was built upon the JNF matrix library, it was quite convenient to manipulate block and submatrices, which is crucial for the Kron reduction. However, verifying the properties of the Kron-reduced Laplacian properties required a detailed block-wise analysis. To address this, we utilized previously verified properties of the Laplacian matrix, such as row-sum and sign constraints, and derived several auxiliary identities involving block matrix–vector equations and block-specific diagonal/off-diagonal entries, which greatly facilitated verification. We also demonstrated a case study on the power dissipation in a resistive electrical network. This example underscored the importance of symmetry to ensure that the total power dissipation is well-defined and physically meaningful. One challenge encountered during this verification was the handling of double summation terms, although scalar–summation multiplication identities appear straightforward, their formal treatment required significant effort in the Isabelle/HOL proof assistant.
\section{Conclusion}
\label{SEC:Conclusion}
\vspace*{-.1cm}
 In this paper, we proposed to use a HOL-based interactive theorem prover to formalize significant network topology matrices. In particular, we formalized the adjacency, degree and Laplacian matrices of weighted directed graphs in Isabelle/HOL and verified their classical properties, including indexing and weight-balanced. Moreover, we formalized the in-incidence and out-incidence matrices and verified several fundamental properties, such as dimensional consistency and indexing.  Subsequently, we established the formal relationships between the incidence matrices, and the adjacency, degree and Laplacian matrices, thereby ensuring the overall correctness and coherence of the formalization. To demonstrate the effectiveness of our formalization, we formally analyzed Kron-reduced Laplacian matrices and verified the power dissipation of a generic resistive electrical network using the Laplacian matrix. As future work, from a formalization point of view, it would be interesting to explore the connection between our formalization and other graph libraries in Isabelle/HOL. Moreover, a promising direction is to extend this formalization toward the analysis of dynamical systems~\cite{bullo2018lectures} across various domains, particularly in verifying their stability and controllability. Another avenue for future work involves developing complex-valued network systems and their corresponding matrix representations, which are extensively employed in graph signal processing~\cite{complex2019graph}.


\bibliographystyle{splncs03_unsrt}
\bibliography{mybib}

\end{document}